\documentclass[fleqn,usenatbib]{mnras}

\usepackage[T1]{fontenc}

\DeclareRobustCommand{\VAN}[3]{#2}
\let\VANthebibliography\thebibliography
\def\thebibliography{\DeclareRobustCommand{\VAN}[3]{##3}\VANthebibliography}

\usepackage{graphicx}	
\usepackage{amsmath}	
\usepackage{amssymb}	

\usepackage{color}

\title[Cluster SZ detection with unbiased noise estimation]{Galaxy cluster SZ detection with unbiased noise estimation: an iterative approach}

\author[\'{I}. Zubeldia et al.]{
\'{I}\~{n}igo Zubeldia,$^{1}$\thanks{E-mail: inigo.zubeldia@manchester.ac.uk}
Aditya Rotti$^{1}$,
Jens Chluba$^{1}$
and Richard Battye$^{1}$
\\
$^{1}$Jodrell Bank Centre for Astrophysics, University of Manchester, Manchester M13 9PL UK
}

\date{Accepted XXX. Received YYY; in original form ZZZ}

\pubyear{2022}

\begin{document}
\label{firstpage}
\pagerange{\pageref{firstpage}--\pageref{lastpage}}
\maketitle

\begin{abstract}
Multi-frequency matched filters (MMFs) are routinely used to detect galaxy clusters from CMB data through the thermal Sunyaev-Zeldovich (tSZ) effect, leading to cluster catalogues that can be used for cosmological inference. In order to be applied, MMFs require knowledge of the cross-frequency power spectra of the noise in the maps. This is typically estimated from the data and taken to be equal to the power spectra of the data, assuming the contribution from the tSZ signal of the detections to be negligible. Using both analytical arguments and \textit{Planck}-like mock observations, we show that doing so causes the MMF noise to be overestimated, inducing a loss of signal-to-noise. Furthermore, the MMF cluster observable (the amplitude $\hat{y}_0$ or the signal-to-noise $q$) does not behave as expected, which can potentially bias cosmological inference. In particular, the observable becomes biased with respect to its theoretical prediction and displays a variance that also differs from its predicted value. We propose an iterative MMF (iMMF) approach designed to mitigate these effects. In this approach, after a first standard MMF step, the noise power spectra are reestimated by masking the detections from the data, delivering an updated iterative cluster catalogue. Applying our iMMF to our \textit{Planck}-like mock observations, we find that the aforementioned effects are completely suppressed. This leads to a signal-to-noise gain relative to the standard MMF, with more significant detections and a higher number of them, and to a cluster observable with the expected theoretical properties, thus eliminating any potential biases in the cosmological constraints.
\end{abstract}

\begin{keywords}
galaxies: clusters: general -- cosmology: observations -- cosmology: diffuse radiation
\end{keywords}



\section{Introduction}


The abundance of galaxy clusters across mass and redshift is a powerful cosmological probe, sensitive to cosmological parameters such as $\Omega_{\mathrm{m}}$ and $\sigma_8$, to the sum of the neutrino masses, and to the nature of the dark sector \citep{Allen2011,Pratt2019}. Cosmic microwave background (CMB) observations provide a unique window into the cosmic galaxy cluster population, allowing for blind cluster detection to high redshifts through the thermal Sunyaev-Zeldovich (tSZ) effect (\citealt{Sunyaev1972}; see \citealt{Carlstrom2002} and \citealt{Mroczkowski2019} for reviews). Over the past decade, a number of galaxy cluster catalogues with $\sim10^2$--$10^3$ objects have been constructed from \textit{Planck}, ACT, and SPT data (e.g., \citealt{Vanderlinde2010,Hasselfield2013,Planck2014,Bleem2015,Planck2016xxvii,Tarrio2019,Aghanim2019,Hilton2020, Melin2021}) and subsequently used in cosmological analyses (e.g., \citealt{Vanderlinde2010,Hasselfield2013,Planck2014XX,Bleem2015,Ade2016,Bocquet2018,Zubeldia2019,Salvati2021}). In coming years, experiments such as the Simons Observatory (SO, \citealt{SO2019}) and CMB-S4 \citep{Abazajian2016} are expected to deliver over an order of magnitude more objects than their predecessors. This will yield unprecedented potential for constraining cosmological parameters, but for it to be realised, systematics across the full analysis pipeline will have to be understood to a level much better than is presently the case. In particular, accurately relating the cluster observable(s) to the cluster mass, which is what is predicted by the theory through the halo mass function, will be crucial for the cosmological success of these upcoming cluster samples \citep{Pratt2019}.

In CMB data, galaxy clusters are typically detected with multi-frequency matched filters (MMFs; \citealt{Herranz2002,Melin2006}), which act directly on CMB frequency maps and exploit the unique cluster spectral and spatial signatures of the tSZ effect. MMFs have indeed been used to produce the main cluster catalogues of \textit{Planck}, ACT, and SPT (see, e.g., \citealt{Bleem2015,Planck2016xxvii,Hilton2020}), and will continue to be used for SO and CMB-S4. For them to be applied in practice, MMFs require knowledge of the cross-channel power spectra (hereafter, `covariance') of the `noise' in the frequency maps, where by the term noise we refer to all the components in the data other than the clusters' tSZ signal itself. This is typically estimated from the data and simply taken to be equal to the data covariance (e.g., \citealt{Planck2016xxvii}), as the tSZ signal is assumed to be small enough to be ignored. So far, however, this assumption has remained to be examined carefully. Given the unprecedented sensitivity to systematics of the upcoming tSZ surveys, it is important that the impact of such assumption on the cluster observables is investigated in detail.

In this work, we study the impact of the presence of the clusters' tSZ signal in the noise covariance estimate on cluster detection with MMFs. We show that, in general, it leads to a loss of signal-to-noise and to a cluster observable, the tSZ amplitude $\hat{y}_0$ or the signal-to-noise $q$, which does not have the expected statistical properties. In particular, we find that $\hat{y}_0$ and $q$ are biased low with respect to their predicted values, a bias that we refer to as `covariance bias'. In addition, $\hat{y}_0$ and $q$ have a standard deviation that is similarly biased low with respect to its prediction. Both effects can potentially lead to a bias in the cosmological inference. We provide a theoretical explanation for these effects, illustrating them with a toy model, and we quantify their magnitude with a set of realistic \textit{Planck}-like cluster catalogues. In order to mitigate them, we propose an iterative MMF (iMMF) approach, in which after a first cluster catalogue is constructed taking the noise covariance to be equal to the data covariance, the detected clusters are {\it masked out} from the data and the noise covariance is then re-estimated. The process can be repeated multiple times until stationary results are obtained. We find our iMMF to be highly effective at suppressing the aforementioned effects, delivering cluster observables that behave as predicted by the theory, as well as leading to a gain in signal-to-noise.

This paper is organised as follows. First, we review the fundamentals of MMF cluster detection in Section \ref{sec:mmf}. Next, in Section \ref{subsec:ps} we address the issue of the contamination of the noise covariance by the cluster signal, deriving analytically the effects it has on the cluster observables and illustrating them numerically with a toy model. We also outline how an iMMF can help in mitigating these effects. We then describe in detail our iMMF implementation in Section \ref{subsec:implementation}, and our mock observations in Section \ref{sec:sims}. Our results are then discussed in Section \ref{sec:results}, where we compare the performance of the standard non-iterative MMF to that of our iMMF, showing the latter to be unaffected by the contamination of the noise covariance by the signal. We finally conclude in Section \ref{sec:conclusion}. In addition, in Appendix \ref{appendix} we derive analytically an expression for the covariance bias of the MMF observables.

\vspace{-4mm}
\section{REVIEW OF MMF FORMALISM}\label{sec:mmf}
Let us consider a set of intensity maps at different frequencies, $\bmath{d}(\bmath{x})$, where the vector dimension of $\bmath{d}$ is equal to the number of frequency channels and where $\mathbfit{x}$ denotes angular position. We assume that  $\mathbfit{d}(\mathbfit{x})$ can be written as
\begin{equation}
    \mathbfit{d} (\mathbfit{x}) = \mathbfit{y} (\mathbfit{x}) + \mathbfit{n} (\mathbfit{x}).
\end{equation}
Here, $\mathbfit{y} (\mathbfit{x})$ is the signal that is being targeted, in our case, the galaxy cluster tSZ signal. We assume that its ith component can be written as $y_i (\mathbfit{x}) = j_{i} (y \ast b_i )(\mathbfit{x})$, where $j_{i}$ is the signal frequency dependency at frequency channel $i$, $y(\mathbfit{x})$ is the Compton-$y$ map, describing the tSZ signal spatial variation, $b_i$ is the instrument beam at frequency channel $i$, and $\ast$ denotes convolution. In addition, $\mathbfit{n} (\mathbfit{x})$ is some additive `noise', which also includes all the other components different from that of interest (the CMB, the kinetic Sunyaev-Zeldovich signal, the Cosmic Infrared Background, Galactic foregrounds, etc.). We further assume that, for a single galaxy cluster, its Compton-$y$ map $y (\mathbfit{x})$ can be written as $y (\mathbfit{x}) = y_0 y_{\mathrm{t}} (\mathbfit{x} ;  \theta_{500}, \bmath{\theta}_\mathrm{c} )$, where $y_0$ is an amplitude parameter, which we take to be the value of the cluster's Compton-$y$ parameter at the cluster centre, and $y_{\mathrm{t}} (\mathbfit{l} ; \theta_{500}, \bmath{\theta}_\mathrm{c})$ is a spatial template. This template depends on the cluster angular size $\theta_{500}$, defined, as is customary, as the angle subtended by $R_{500}$, the radius within which the mean density is equal to 500 times the critical density at the cluster's redshift. It also depends on the sky coordinates of its centre, $\bmath{\theta}_\mathrm{c}$. The tSZ signal template at frequency $i$ can then be written as $y_{\mathrm{t},i} (\mathbfit{x};  \theta_{500}, \bmath{\theta}_\mathrm{c} ) = j_{i} (y_{\mathrm{t}} \ast b_i )(\mathbfit{x};  \theta_{500}, \bmath{\theta}_\mathrm{c} )$. At some given input values of $\theta_{500}$ and $\bmath{\theta}_\mathrm{c}$, an estimator for $y_0$ can then be written, assuming the flat-sky approximation, as \citep{Melin2006}
\begin{equation}\label{estimator}
        \hat{y}_0 =  N^{-1} \int  \frac{ d^2 \mathbfit{l}}{2 \pi}  \,  \mathbfit{y}_{\mathrm{t}}^{\dagger} (\mathbfit{l})   \mathbfss{N}^{-1} (l) \mathbfit{d} (\mathbfit{l}),
\end{equation}
where the dagger denotes conjugate transposition, $\mathbfss{N} (l)$ is the \textit{noise} cross-channel power spectrum matrix, and the normalisation $N$ is given by
\begin{equation}\label{noise}
   N =   \int  \frac{ d^2 \mathbfit{l}}{2 \pi} \, \mathbfit{y}_{\mathrm{t}}^{\dagger} (\mathbfit{l}) \mathbfss{N}^{-1} (l) \mathbfit{y}_{\mathrm{t}} (\mathbfit{l}).
\end{equation}
Here, we have dropped the dependency of the template, and hence of the estimator, on $\theta_{500}$ and $\bmath{\theta}_\mathrm{c}$ to avoid clutter in the notation. This estimator is known as a multi-frequency matched filter estimator (see, e.g., \citealt{Melin2006}), so-called because it acts directly on the multi-frequency data. Its variance is given by $\sigma^2_{y_0} = N^{-1}$, and we note that it is unbiased only if the input $\theta_{500}$ is equal to its true value and if the template is placed at the cluster's true sky location (for a more detailed discussion, see \citealt{Zubeldia2021}). 

The signal-to-noise of the detection, $q$, can be written as
\begin{equation}\label{snr}
    q = \frac{\hat{y}_0}{\sigma_{y_0}} = N^{-1/2} \int  \frac{ d^2 \mathbfit{l}}{2 \pi}  \,  \mathbfit{y}_{\mathrm{t}}^{\dagger} (\mathbfit{l})   \mathbfss{N}^{-1} (l) \mathbfit{d} (\mathbfit{l}).
\end{equation}
At fixed input parameters $\theta_{500}$ and  $\bmath{\theta}_\mathrm{c}$, and assuming $\mathbfss{N} (l)$ to be noise-free, $q$ is an unbiased estimator of the true signal-to-noise $\bar{q}_{\mathrm{t}}$, which is given by
\begin{equation}\label{truesnr}
    \bar{q}_{\mathrm{t}} = \frac{y_0}{\sigma_{y_0}} = N^{-1/2}  \int  \frac{ d^2 \mathbfit{l}}{2 \pi}  \,  \mathbfit{y}_{\mathrm{t}}^{\dagger} (\mathbfit{l})   \mathbfss{N}^{-1} (l) \mathbfit{y} (\mathbfit{l}),
\end{equation}
where, we recall, $\mathbfit{y}$ is the Compton-$y$ signal. In addition, the estimator $q$ has unity standard deviation by construction. We note that, typically, galaxy clusters are detected by maximising the signal-to-noise $q$ over a set of parameters, often $\theta_{500}$ and $\bmath{\theta}_\mathrm{c}$, a procedure that leads to an \textit{optimal} signal-to-noise for each detection, which we denote by $q_{\mathrm{opt}}$. This is usually implemented in practice by producing a set of signal-to-noise maps for a range of input values of $\theta_{500}$. In each map, the value of a given pixel corresponds to the MMF signal-to-noise given by placing the template at the pixel centre. Clusters are then identified as the peaks in these signal-to-noise maps above some threshold. As shown in \citet{Zubeldia2021}, for values of $q_{\mathrm{opt}}$ of about 5 and above, this optimal signal-to-noise $q_{\mathrm{opt}}$ follows approximately a unit-variance Gaussian with mean equal to $(\bar{q} + f)^{1/2}$, where $f$ is the number of parameters that the signal-to-noise is being maximised for, usually 3 (angular size and sky location).

\vspace{-4mm}
\section{MMF noise power spectra estimation}\label{subsec:ps}
In order for a MMF to be used in practice, the noise cross-channel power spectrum matrix (hereafter, the `noise covariance'), $\mathbfss{N} (l)$ in, e.g., Eq. (\ref{estimator}), must be determined. This is typically estimated from the data and taken to be equal to the data cross-channel power spectrum matrix (hereafter, the `data covariance'), which we denote with $\mathbfss{C} (l)$. Indeed, the tSZ signal itself is often claimed to be small enough to be safely ignored. 

This procedure, however, introduces two potentially problematic effects. First, the clusters' tSZ signal may in fact not be negligible, especially for high-significance detections, leading to an overestimation of the noise covariance. In addition, the noise covariance becomes dependent on the data, i.e., noisy, making the MMF estimator, $\hat{y}_0$ or $q$, no longer linear in the data, and hence potentially introducing a bias. We note that this second point is true in general for any data-based covariance, regardless of whether it contains the signal or not. In this section we consider the impact of these two effects on the MMF mean and variance.

\subsection{Impact on the MMF variance}\label{sec:noise}

Let us first consider the impact of covariance estimation on the MMF variance. If the data covariance $\mathbfss{C} (l)$ is used instead of the noise covariance $\mathbfss{N} (l)$, the MMF estimate of $y_0$ is given by
\begin{equation}
        \hat{y}_0 [\mathbfss{C}] =  N^{-1} [\mathbfss{C}] \int  \frac{ d^2 \mathbfit{l}}{2 \pi}  \,  \mathbfit{y}_{\mathrm{t}}^{\dagger} (\mathbfit{l})   \mathbfss{C}^{-1} (l) \mathbfit{d} (\mathbfit{l}),
\end{equation}
where we have made explicit its dependence on $\mathbfss{C}$, and where $N^{-1} [\mathbfss{C}]$ is given by Eq. (\ref{noise}) by substituting $\mathbfss{N}$ for $\mathbfss{C}$. As noted above, in general $\mathbfss{C} (l)$ depends on the data, making $\hat{y}_0$ a non-linear function of $\mathbfit{d} (\mathbfit{l})$. However, to lowest order in the data, i.e., neglecting the noisy character of  $\mathbfss{C} (l)$, it is linear. Here, we will only consider the estimator variance to this order, neglecting higher-order terms. The variance of $\hat{y}_0 [\mathbfss{C}]$ to linear order in the data then is given by
\begin{equation}\label{truestd}
    \sigma_{y_0}^2 [\mathbfss{C}] = N^{-2} [\mathbfss{C}]  \int  \frac{ d^2 \mathbfit{l}}{2 \pi}  \, \mathbfit{y}_{\mathrm{t}}^{\dagger}  (\mathbfit{l}) \mathbfss{C}^{-1} (l) \mathbfss{N} (l)  \mathbfss{C}^{-1} (l) \mathbfit{y}_{\mathrm{t}}  (\mathbfit{l}).
\end{equation}
This is larger than the variance of the optimally-weighted estimator, $\sigma_{y_0}^2 [\mathbfss{N}]$, as optimality is achieved when the true noise covariance is used. Therefore, the estimator is suboptimal. In addition, $\sigma_{y_0}^2 [\mathbfss{C}]$ is smaller than the variance that would be naively predicted assuming $\mathbfss{C}$ to be the noise covariance, $\sigma_{y_0,\mathrm{p}}^2 [\mathbfss{C}] = N^{-1} [\mathbfss{C}]$. Indeed, we can write

\begin{equation}\label{truestd}
    \sigma_{y_0}^2 [\mathbfss{C}] = \sigma_{y_0,\mathrm{p}}^2 [\mathbfss{C}] N^{-1} [\mathbfss{C}]  \int  \frac{ d^2 \mathbfit{l}}{2 \pi}  \, \mathbfit{y}_{\mathrm{t}}^{\dagger}  (\mathbfit{l}) \mathbfss{C}^{-1} (l) \mathbfss{N} (l)  \mathbfss{C}^{-1} (l) \mathbfit{y}_{\mathrm{t}}  (\mathbfit{l}) .
\end{equation}
If $\mathbfss{C} (\mathbfit{l}) > \mathbfss{N} (\mathbfit{l})$ for all $\mathbfit{l}$, which is always the case if the noise and the signal are uncorrelated, the factor multiplying $\sigma_{y_0,\mathrm{p}}^2 [\mathbfss{C}]$ is less than unity, and hence $ \sigma_{y_0}^2 [\mathbfss{C}] < \sigma_{y_0,\mathrm{p}}^2 [\mathbfss{C}]$. Thus, if $N^{-1} [\mathbfss{C}]$ is used as the associated variance in a likelihood analysis of a set of measured values of $y_0$, the uncertainty on $y_0$ will be overestimated. In short, if the data covariance is used as the noise covariance, there are two different sources of suboptimality: first, the data is filtered in a suboptimal way, leading to noisier matched filter measurements than in the optimal case. In addition, the associated uncertainty on these measurements is overestimated. This will lead to a degradation of the constraining power of a likelihood (e.g., number counts) analysis with respect to the optimal case, and the latter effect may also lead to a bias, as it constitutes an error in the modelling of the observable.



Let us now consider the signal-to-noise. If $\mathbfss{C}$ is used in the matched filter, the signal-to-noise of a detection is given by
\begin{equation}
    q [\mathbfss{C}] = \frac{\hat{y}_0 [\mathbfss{C}]}{\sigma_{y_0,\mathrm{p}} [\mathbfss{C}]} = N^{-1/2} [\mathbfss{C}] \int  \frac{ d^2 \mathbfit{l}}{2 \pi} \mathbfit{y}_{\mathrm{t}}^{\dagger}  \mathbfss{C}^{-1} (l) \mathbfit{d} (\mathbfit{l}).
\end{equation}
Since the predicted standard deviation of $\hat{y}_0$, $N^{-1/2} [\mathbfss{C}]$, is larger than the optimal one, $N^{-1/2} [\mathbfss{N}]$, the detection signal-to-noise is smaller than in the optimal case. (Note that if the real standard deviation, $\sigma_{y_0} [\mathbfss{C}]$, was used instead, the signal-to-noise would also be lower than in the optimal case: the loss of signal-to-noise is real.) In addition, since the predicted standard deviation of $\hat{y}_0$ is greater than its true standard deviation, given by Eq. (\ref{truestd}), the real standard deviation of the signal-to-noise is smaller than unity. The variance of $q$ can indeed be written as
\begin{equation}\label{qstd}
    \sigma_{q}^2 [\mathbfss{C}] = N^{-1} [\mathbfss{C}]  \int  \frac{ d^2 \mathbfit{l}}{2 \pi}  \, \mathbfit{y}_{\mathrm{t}}^{\dagger}  (\mathbfit{l}) \mathbfss{C}^{-1} (l) \mathbfss{N} (l)  \mathbfss{C}^{-1} (l) \mathbfit{y}_{\mathrm{t}}  (\mathbfit{l}),
\end{equation}
which, if $\mathbfss{C} (\mathbfit{l}) > \mathbfss{N} (\mathbfit{l})$ for all $\mathbfit{l}$, is less than unity. Assuming $\sigma_q$ to be unity in a likelihood analysis would lead to further loss of constraining power, this effect being equivalent to the second source of suboptimality noted above, and could also lead to a bias, as it is an error in the modelling of the observable.


\subsection{Impact on the MMF mean}\label{sec:bias}
Let us now consider the impact of covariance (mis)estimation on the MMF mean. At fixed input parameters (i.e., $\theta_{500}$ and sky location), $\hat{y}_0$ and $q$ are given by Eq. (\ref{estimator}) and Eq. (\ref{snr}), respectively. It is clear that, for any choice of `covariance' $\mathbfss{N} (l)$ that is independent from the data vector $\mathbfit{d}$, $\hat{y}_0$ and $q$ are unbiased estimators of $y_0$ and the true signal-to-noise, respectively. That is, $\left\langle \hat{y}_0 \right\rangle = y_0$ and $\left\langle q \right\rangle = \bar{q}_{\mathrm{t}}$, where $\bar{q}_{\mathrm{t}}$, the true signal-to-noise, is given by Eq. (\ref{truesnr}), and where angular brackets denote ensemble averaging over `noise' $\mathbfit{n}$. Thus, if we think of either estimator as a function of the data $\mathbfit{d}$, it is unbiased to lowest, i.e., linear, order in $\mathbfit{d}$ even if the wrong covariance is used. This is unlike the matched filter variance, which is biased to lowest order in the data, as shown in the previous section. %

Let us now consider the next-order term if an empirically-estimated, i.e., noisy, data covariance, $\mathbfss{C} (l; \mathbfit{d})$, is used. This term is linear in the `covariance perturbation' $[\mathbfss{C} (l; \mathbfit{d}) -   \mathbfss{C}_0 (l)] / \mathbfss{C}_0 (l)$, where $\mathbfss{C}_0$ is the `true', noiseless data covariance, and hence third-order in the data. Let us consider the signal-to-noise $q$. In Appendix \ref{appendix} we show that, to this order in the covariance perturbation and neglecting the noise bispectrum, $ \left\langle q - \bar{q}_{\mathrm{t}} \right \rangle$ is given by
\begin{equation}\label{biasprediction}
    \left\langle q - \bar{q}_{\mathrm{t}} \right \rangle = \Delta \bar{q}_1 + \Delta \bar{q}_2,
\end{equation}
where $\Delta \bar{q}_1$ and $\Delta \bar{q}_2$  are given by Eqs. (\ref{bias1}) and (\ref{bias2}), respectively, both in Appendix \ref{appendix}. Both $\Delta \bar{q}_1$ and $\Delta \bar{q}_2$ are non-zero if the signal is present in the covariance estimate and zero otherwise (assuming that the noise bispectrum can be neglected). Thus, to first order in the covariance perturbation, $q$ is a biased estimator of the true signal-to-noise $\bar{q}_{\mathrm{t}}$ if the covariance estimate is contaminated by the signal. We refer to this bias as the matched filter covariance bias. A similar expression can be found for $\hat{y}_0$, which is also biased to this order (see Appendix \ref{appendix}). This can be a significant bias, as we illustrate in the next section and in Section \ref{sec:results}, which can potentially lead to biased cosmological inference. We note that this bias is present even if the matched filter template matches perfectly the real cluster profile: it is entirely due to the presence of the signal in the covariance estimate. On the other hand, if the covariance estimate does not include the signal, and if the bispectrum of the noise can be neglected, $q$ remains unbiased to this order in the covariance perturbation. Hence, any method that successfully removes the detections in the covariance estimation process will mitigate against the covariance bias.

We note that $ \left\langle q - \bar{q}_{\mathrm{t}} \right \rangle$ is the relevant quantity to consider, and not $ \left\langle q \right\rangle$. Indeed, to this order in the covariance perturbation, $\bar{q}_{\mathrm{t}}$ also depends on the data via the covariance. Calculating $ \left\langle q \right\rangle$ would yield the bias of $q$ with respect to a true signal-to-noise defined with the true, noiseless covariance. However, this true signal-to-noise is not used in practice. Indeed, in the modelling of the observable $q$ in, e.g., a likelihood, it is the true signal-to-noise computed with the noisy covariance that typically appears. Thus, to linear or higher order in the covariance perturbation, we define the true signal-to-noise as computed with the noisy covariance. The bias of $q$ with respect to this true signal-to-noise is then given by $ \left\langle q - \bar{q}_{\mathrm{t}} \right \rangle$, which is therefore the relevant quantity to consider.

\subsection{Illustration in an idealised scenario}\label{sec:toymodel}
In order to illustrate the impact of covariance misestimation on the matched filter bias and variance, and to get order-of-magnitude estimates, we produce an idealised set of mock observations. More specifically, we consider single-frequency observations of individual clusters, with a square field of $256 \times 256$ pixels for each cluster, centred at the cluster centre and with a pixel size of 1\,arcmin and a Gaussian beam with a FWHM of 5\,arcmin. We consider a set of six cluster masses linearly spaced between $M_{500} = 3 \times 10^{14} M_{\odot}$ and $M_{500} = 6 \times 10^{14} M_{\odot}$ and a single redshift $z=0.25$. We produce $10^3$ Compton-$y$ maps for each cluster mass assuming the pressure profile of \citet{Arnaud2010}, adding white noise with a pixel standard deviation of $10^{-5}$. For each map, we make two estimates of the noise covariance, one by computing the power spectrum (`covariance') of the total data (cluster signal + noise) map (`biased covariance'), and the other one by applying the same procedure to the noise map alone (`unbiased covariance'). We average each covariance estimate in a set of eight bins linearly spaced between $l=0$ and $l=2700$. Then, we use these two covariance estimates to construct two matched filter estimators, which we evaluate at the true cluster parameters (true angular size and sky location), making two signal-to-noise measurements for each cluster. We also compute two values for the true signal-to-noise $\bar{q}_{\mathrm{t}}$ for each cluster, obtained by matched filtering the signal-only map with the `unbiased covariance' and `biased covariance' matched filters, respectively.

\begin{figure}
\centering
\includegraphics[width=0.5\textwidth,trim={1mm 12mm 1mm 18mm},clip]{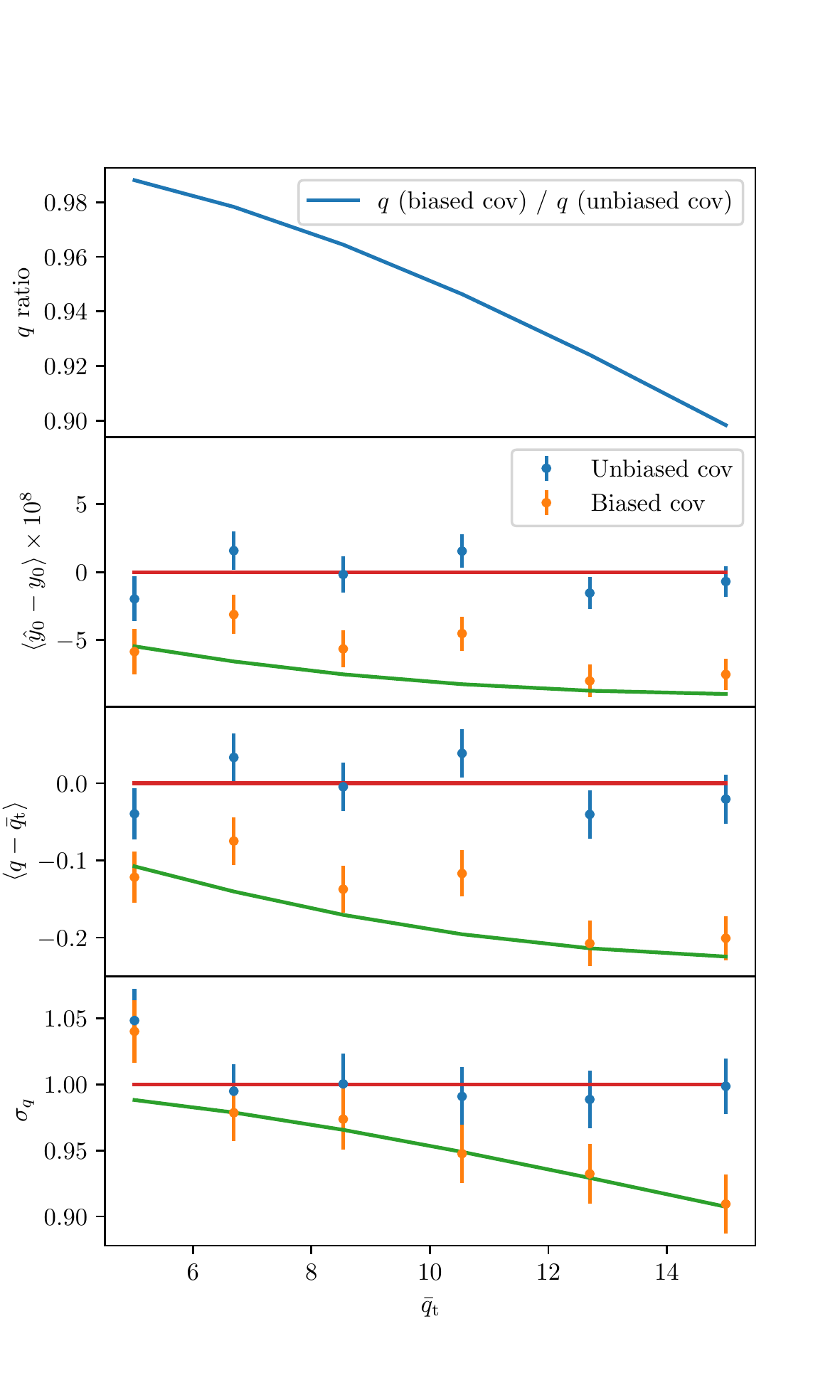}
\\
\caption{Results from the simple numerical experiment of Section~\ref{sec:toymodel}. \textit{First panel}: the ratio of the biased covariance true signal-to-noise to the unbiased covariance true signal-to-noise. \textit{Second and third panel}: empirical mean of the $\hat{y}_0$ and $q$ residuals as a function of the true signal-to-noise $\bar{q}_{\mathrm{t}}$, for the biased covariance (blue points) and the unbiased covariance (orange points). The theoretical prediction from Eq.~\eqref{biasprediction} is shown as the green curve. \textit{Fourth panel}: Empirical standard deviation of the signal-to-noise residuals, with the theoretical prediction from Eq.~\eqref{qstd} in green.}
\label{fig:bias_toy}
\end{figure}

The first panel of Figure \ref{fig:bias_toy} shows the ratio of the biased covariance true signal-to-noise to the unbiased covariance true signal-to-noise, as a function of the unbiased covariance true signal-to-noise. The signal-to-noise loss due to the presence of the signal in the covariance is clear, with it increasing with signal-to-noise. The third panel depicts the empirical mean of the signal-to-noise residuals, $q - \bar{q}_{\mathrm{t}}$, as a function of the true signal-to-noise for both matched filters (blue points for the unbiased covariance and orange points for the biased covariance), along with the theoretical prediction of the bias given by Eq. (\ref{biasprediction}) (green curve). No bias is seen for the unbiased covariance matched filter, whereas a $\simeq 10$\,\% negative bias can be seen in the biased covariance case, which increases with signal-to-noise. 
The second panel is an analogous plot for the $\hat{y}_0$ estimator. As with $q$, the estimator is unbiased for the unbiased covariance case, but a negative bias appears if the biased covariance containing the signal is used instead. The existence of both biases validates numerically the points made in Section~\ref{sec:bias}. We note that our theoretical predictions for their value, given by Eq. (\ref{biasprediction}) for $q$ and by an analogous equation for $\hat{y}_0$ (see Appendix~\ref{appendix}), provide a rather good fit to the observed values. Finally, the fourth panel shows the empirical standard deviation of the signal-to-noise measurements as a function of the true signal-to-noise. If the covariance is unbiased, the standard deviation is consistent with unity throughout (blue points), as expected. If, however, it includes the signal, the standard deviation is less than one, also as expected, being instead consistent with the prediction given by Eq. (\ref{qstd}), which is shown as a green curve. This is all in agreement with the description of this effect in Section~\ref{sec:noise}.

We note that the magnitude of the covariance bias, as well as of the deviation of the standard deviation from unity, depends on observational specifications such as the field size relative to the angular extent of the signal and the binning used for the estimation of the covariance. Quantifying these dependencies in any detail is beyond the scope of this paper.

\subsection{Unbiased noise power spectra: masking detected clusters}\label{subsec:itexplain}
It is clear, both from the theoretical considerations of Sections \ref{sec:noise} and \ref{sec:bias} and from the numerical experiment of Section \ref{sec:toymodel}, that the presence of the signal in the estimate of noise covariance leads to a loss of signal-to-noise and to biases in the modelling of the observable. A possible approach to avoid these issues involves masking the detections above some signal-to-noise threshold from the data maps, and then reestimating the noise covariance with the masked maps, producing a new set of detections. This step can be iterated until no new detections above the masking threshold appear. This constitutes an \textit{iterative} approach, which refer to as iterative multi-frequency matched filter, or iMMF.

\begin{figure}
\centering
\includegraphics[width=0.5\textwidth,trim={1mm 2mm 1mm 8mm},clip]{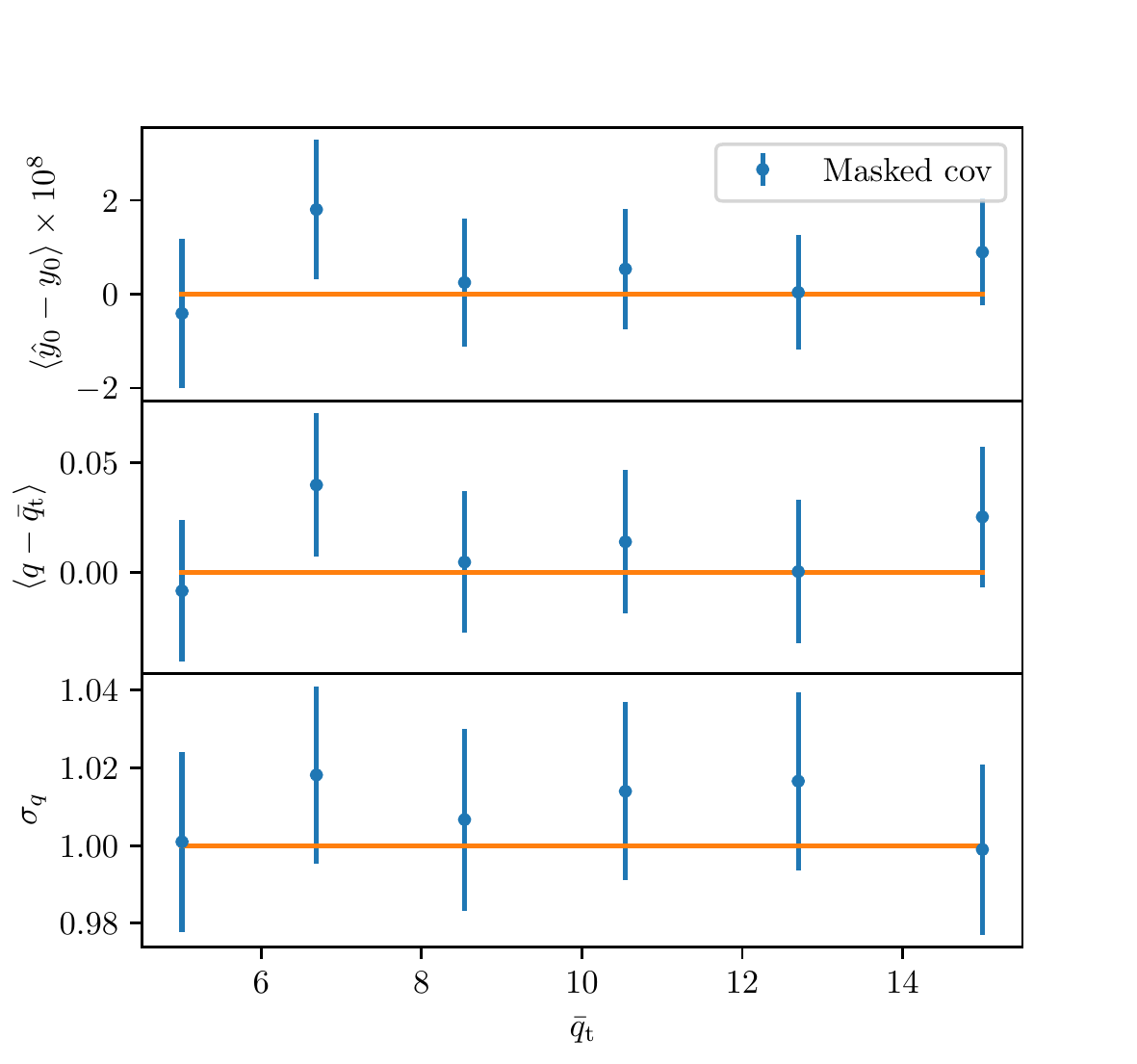}
\caption{Results from the numerical experiment of Section \ref{sec:toymodel} when the cluster is masked for covariance estimation. The three panels are analogous to the three lower panels of Figure \ref{fig:bias_toy}. There is no obvious bias in these idealised mock observations.}
\label{fig:bias_toy_masked}
\end{figure}

As a simple illustration of this approach, we produce iterative-like measurements for our mock observations of Section \ref{sec:toymodel}. Namely, we estimate the matched filter covariance by simply masking the central cluster with a circular mask with a radius equal to $3 \theta_{500}$, accounting for the missing sky fraction by rescaling the covariance by $1/f_{\mathrm{sky}}$. Our results are shown in Figure \ref{fig:bias_toy_masked}, which is analogous to Figure \ref{fig:bias_toy} for the iterative case. No bias is observed and the correct standard deviation is returned. We note that we do not plot the iterative true signal-to-noise against the unbiased true signal-to-noise because they are essentially the same, agreeing to less than 0.1\,\%. In Section \ref{sec:results} we present analogous results for a more realistic observational setting, with the more realistic implementation of the iMMF used to obtain them being described in Section \ref{subsec:implementation} and the production of the simulated observations in Section \ref{sec:sims}.

We note that our iterative approach has the advantage of completely removing the tSZ signal of the detections for noise estimation purposes in a clean, model-agnostic way. An alternative possibility could involve iteratively subtracting a cluster model from the frequency maps for each of the detections above some threshold, as proposed in \citet{Zubeldia2021}, instead of masking them. This approach, however, has the disadvantage of being model-dependent, removing the signal in as much as the model is a good description of cluster profile. In addition, even if the model is perfectly matched to the real profile, residuals always remain and convergence is not guaranteed, as we have verified numerically. Indeed, the model parameter estimates used to evaluate the model are always noisy. If the estimate of the amplitude, $\hat{y}_0$, fluctuates up significantly with respect to its true value, the cluster signal can be overestimated, and hence the noise covariance can be underestimated. This can lead to a further overestimation of $\hat{y}_0$ in the next iteration, which in turn can lead to an even more underestimated noise covariance. Eventually, grossly underestimating the noise matrix could lead to many false detections. Hence, due to these issues, we favour our `masking' iterative approach over this `subtraction' approach.

\section{Our MMF implementation}\label{subsec:implementation}
In this section, we describe in detail our implementation of the iterative MMF (iMMF) cluster finding method. This implementation takes as input a set of intensity maps at different frequencies defined on the sphere, as well as a Galactic and a point source mask, and produces two cluster candidate catalogues, an iterative one and a non-iterative one, with estimated values for the signal-to-noise, the cluster angular scale $\theta_{500}$, and the cluster sky coordinates. It can also provide values for the clusters' true signal-to-noise if true tSZ maps are provided. In addition, our implementation can operate in a `fixed' mode in which the MMF is evaluated at an input set of model parameters, $\theta_{500}$ and sky location. Our implementation can be applied to both simulated and real data.


\subsection{Sky tessellation}\label{subsec:tesselation}

Our implementation takes as input a set of intensity maps defined on the full sky, discretised according to the HEALPix pixelisation scheme \citep{Gorski2005}. In this work, we consider maps with $N_{\mathrm{side}} = 2048$, which is the same pixelisation that was used in the \textit{Planck} HFI legacy maps. Our implementation also takes two binary full-sky masks, similarly defined in the HEALPix scheme: a Galactic mask zeroing regions deemed to be too contaminated by Galactic emission, and a point-source mask. As instrumental noise and foregrounds can vary significantly across the sky, the sky is tessellated into a set of tiles, each of which are analysed independently. As well as allowing for local noise estimation, this also enables for the application of the flat-sky formalism described in Section~\ref{sec:mmf}, ensuring that in any tile there are no points too far away from the projection centre. 

The sky tessellation that we choose is that provided by a HEALPix pixelisation with $N_{\mathrm{side}}=8$, which yields a total of 768 tiles, each with an area of 53.7\,deg$^2$ and a typical extent of $53.7^{1/2} $= 7.3\,deg. This tessellation scheme guarantees that the entirety of the sky is covered and that each point in the sky belongs to only one tile, both of which are desirable properties. Indeed, the former means that the survey area is maximised (in practice, to the extent allowed by the masks), whereas the latter implies that the survey completeness, i.e., the probability of detection at given true cluster parameters, is well-defined for each tile. This is unlike other tessellation schemes used in previous work (e.g., in \citealt{Planck2016xxvii}), in which square or rectangular tiles are used and, as a consequence, some areas of the sky are included in more than one tile, leading to a less well-defined completeness.

Once the sky is tessellated, square cut-outs (hereafter, `fields') of 14.8\,deg $\times$ 14.8\,deg are extracted as equirectangular (or \textit{plate carr\'{e}e}) projections centred at each tile centre. Each field consists of 1024 $\times$ 1024 pixels, with a pixel size of 0.867\,arcmin. Such fields are produced for all the frequency maps (`frequency fields'), as well as for the two masks (`mask fields'). The chosen cut-out size ensures that every tile is fully contained within its corresponding field, regardless of its shape. In order to suppress spectral leakage when taking Fourier transforms, each of the Galactic mask fields are apodised with the `smooth' scheme of \texttt{pymaster}\footnote{\texttt{namaster.readthedocs.io}}, which zeroes all pixels closer to 2.5 times the apodisation scale to a masked pixel, then convolves the resulting mask with a Gaussian kernel with its standard deviation equal to the apodisation scale, and finally zeroes all the originally-masked pixels. We choose an apodisation scale of 12\,arcmin, noting the edges of the mask fields are also apodised by this procedure.

\subsection{Field preprocessing}\label{sec:preprocess}

Some preprocessing of the the frequency fields is required before the MMF is applied on them. First, the point source mask is applied on each field. The masked regions are then inpainted with the real-space diffusive inpainting algorithm of \citet{Gruetjen2015}, using a total 100 iterations, after which the apodised Galactic mask is applied. Then, a custom Fourier-space filter can be applied. For this work, we use a Fourier-space top-hat with a minimum multipole $l_{\mathrm{min}} = 100$ and a maximum multipole $l_{\mathrm{max}} = 2500$. 

\subsection{Noise covariance estimation}\label{subsubsec:noise}

In order for the MMF to be applied, the noise cross-channel power spectrum matrix (`noise covariance') must be estimated. For each field, the cross-channel power spectra of the inpainted, apodised frequency fields are computed using the MASTER algorithm \citep{Hivon2002} as implemented in \texttt{pymaster} \citep{Alonso2019}. The cross power spectra are estimated in a set of 255 equally-spaced bands centred between $l = 48.6$ and $l = 12405$, properly accounting for the multiple coupling due to the Galactic mask by multiplying each pseudo spectrum by the inverse of the mask coupling matrix. 

In the first iteration of the iterative MMF, the noise covariance is computed with the input frequency maps. As we detail below, in the next iterations, the same noise covariance estimation pipeline is run on frequency maps in which the detections above some signal-to-noise threshold are masked out.

\subsection{MMF: production of signal-to-noise maps}
Once the noise covariance is computed, for each tile, a set of MMFs is constructed. For the template $\mathbfit{y}_{\mathrm{t}} (\bmath{x})$ we use the Compton-$y$ map due to the \citet{Arnaud2010} pressure profile, which, for each channel, is multiplied by the non-relativistic tSZ spectral energy distribution and convolved by the corresponding instrument beam and HEALPix pixel transfer function. We set the concentration to be $c=1.177$ following the \textit{Planck} analyses (e.g., \citealt{Planck2016xxvii}). The pressure profile is integrated numerically up to a radius of $5 R_{500}$, at which it is truncated. In the `cluster finding' mode, in which clusters are blindly detected, this template is evaluated at a set of 15 values of $\theta_{500}$ logarithmically spaced between $\theta_{500}=0.5$\,arcmin and  $\theta_{500}=15$\,arcmin, yielding a total of 15 matched filters. On the other hand, in the `fixed mode', the matched filter is evaluated at the values of $\theta_{500}$ of the clusters of the input catalogue falling within such tile.

The matched filtering operation is implemented as a sum of convolutions of two vector fields: (1) the (inpainted, apodised) frequency fields, $\mathbfit{d} (\mathbfit{l})$, and (2), the beam-convolved inverse-variance weighted template, which in Fourier space reads
\begin{equation}
  \mathbfit{y}_{\mathrm{t},i}^{\mathrm{f}} ( \mathbfit{l} ;  \theta_{500}) = \mathbfss{N}^{-1} (l) \mathbfit{y}_{\mathrm{t}} (\mathbfit{l};  \theta_{500}),
\end{equation}
where $\mathbfss{N} (l)$ is the noise covariance estimate and $ \mathbfit{y}_{\mathrm{t}}$ is the beam-convolved tSZ template at frequency $i$ and input angular scale $\theta_{500}$, $y_{\mathrm{t},i} (\mathbfit{x};  \theta_{500}) = j_{i} (y_{\mathrm{t}} \ast b_i )(\mathbfit{x};  \theta_{500})$. Here, we recall that $j_{i}$ and $b_i$ are, respectively, the tSZ frequency dependence and the instrument beam at frequency channel $i$. We note that the same Fourier-space filter that was applied to the frequency fields is also applied to the template. The sum of the frequency-wise convolutions of these two fields normalised by $N^{-1}$, where $N^{-1}$ is given by Eq.~\eqref{noise}, produces a signal-to-noise map at the input filter angular scale $\theta_{500}$,
\begin{equation}\label{qmap}
    q (\mathbfit{x} ;  \theta_{500}) = N^{-1/2} \sum_i ( y_{\mathrm{t},i}^{\mathrm{f}} \ast d_i) (\mathbfit{x}).
\end{equation}
This signal-to-noise map corresponds to the matched filter signal-to-noise, as given in Eq. (\ref{snr}), evaluated at the input angular scale $\theta_{500}$ and at the centre of each pixel of the input frequency fields. Implementing this operation as a sum of convolutions significantly speeds up its computation. We note that, for each $q$ map, a map of the estimated central Compton-$y$ parameter is also produced. This is simply given by Eq. (\ref{qmap}), substituting $N^{-1/2}$ by $N^{-1}$.

In the fixed mode, the values of $q$ and $\hat{y}_0$ for each cluster in the input catalogue are directly extracted from the corresponding maps at the input sky locations. We note that our implementation supports subgrid extraction, in which, for each cluster, the corresponding MMF is centred precisely in the actual sky location, and not, e.g., in the closest pixel centre. If the template matches the real cluster profile perfectly, this yields unbiased matched filter measurements of $q$ and $y_0$. Indeed, placing the template at the closest pixel centre would lead to a miscentering bias, with $q$ and $\hat{y}_0$ being, in general, biased low with respect to their true values. We also note that if an iterative covariance is desired in the fixed mode, a cluster-finding step is performed first, so that the detections can be masked. This cluster-finding step is done in exactly the same way as in the cluster finding mode.

\subsection{Peak finding}\label{subsec:peaks}
In the cluster finding mode, the final step is to turn the signal-to-noise maps into a catalogue of cluster candidates. First, for a given tile, the signal-to-noise maps are masked with a `selection mask', which is defined as the product of two masks. The first one is a product of the unapodised Galactic mask with the point source mask, increased by zeroing all the pixels closer to masked pixels by a buffer distance of 10\,arcmin. This buffering procedure minimises the chances of spurious detections close to the mask edges due to Fourier ringing. The second mask is the tesselation mask, a binary mask zeroing all the pixels falling outside the corresponding HEALPix tile, which ensures that each point in the sky is only covered once for the purposes of cluster detection. 

A first set of cluster candidates is then formed as the local peaks of the three-dimensional masked signal-to-noise field $q (\mathbfit{x} ;  \theta_{500})$ that are above a selection threshold $q_{\mathrm{th}}$, which we choose to be $q_{\mathrm{th}} = 4$. These local peaks are found with a real-space peak-finding filter, which identifies a peak as a (three-dimensional) pixel with a signal-to-noise value larger than any of its neighbouring pixels. For each detection, the peak signal-to-noise value is assigned as the value of its the optimal signal-to-noise $q_{\mathrm{opt}}$. In addition, the corresponding values of $\hat{y}_0$ and of the filter angular scale $\theta_{500}$ are assigned as the matched estimates for $y_0$ and $\theta_{500}$, and the pixel coordinates of the peak as the sky coordinates of the detection. As was detailed in \citet{Zubeldia2021}, these estimates of $y_0$, $\theta_{500}$ and sky location can be seen as maximum-likelihood estimates, and as such are, in general, biased (except the two sky coordinates, due to the symmetry of the problem).

\subsection{Iterative noise covariance estimation}
The catalogue of clusters candidates produced by the method described thus far is constructed assuming that the noise covariance matrix is equal to the data covariance matrix. Our implementation offers the possibility of stopping here, yielding a non-iterative cluster catalogue. However, this first cluster catalogue can also be seen as the product of the first step of the iterative MMF (iMMF) algorithm. Here we describe how a refined, iterative catalogue is produced.

First, a binary `cluster mask' is produced by zeroing all the pixels falling within $3 \hat{\theta}_{500}$\,arcmin of the estimated centre of each cluster candidate with $q_{\mathrm{opt}}$ above a masking threshold. We choose this threshold to be $q_{\mathrm{mask}} = 4$, i.e., equal to the detection threshold. This is to ensure that the covariance bias for the detected clusters vanishes. Indeed, as shown in Section \ref{sec:bias}, the covariance bias is caused, for each detected cluster, by the presence of that cluster itself in the covariance estimate. We note that in the construction of the cluster mask we also include cluster candidates which fall within the buffered Galactic + point source mask, but outside the tessellation mask. We do so because they also contribute to the noise covariance estimate, as the tessellation mask is only imposed in the peak-finding step. 

Once the cluster mask is computed, it is multiplied by the point source mask, leading to a new `point source' mask. The pipeline is then re-run from the preprocessing step of Section \ref{sec:preprocess} using the new cluster + point source mask instead of the original point source mask in the covariance estimation step. This produces a new noise covariance estimate, which is then used to construct a new set of MMFs, which in turn lead to an updated cluster catalogue. We note that the original selection mask (buffered Galactic + point source mask times the tessellation mask) is used in the peak-finding step: the cluster candidates are only masked out for noise estimation purposes. Indeed, otherwise they would not be detected in this second iteration.

In this work we apply only one iteration of the method, which in Section \ref{sec:results} we show to be enough to mitigate the effects of covariance contamination by the signal. However, our iMMF implementation also allows for more iterations, with the stopping criterion that no new detections above the masking threshold are found.

\begin{figure*}
\centering
\includegraphics[width=\textwidth, trim={5mm 14mm 5mm 10mm},clip]{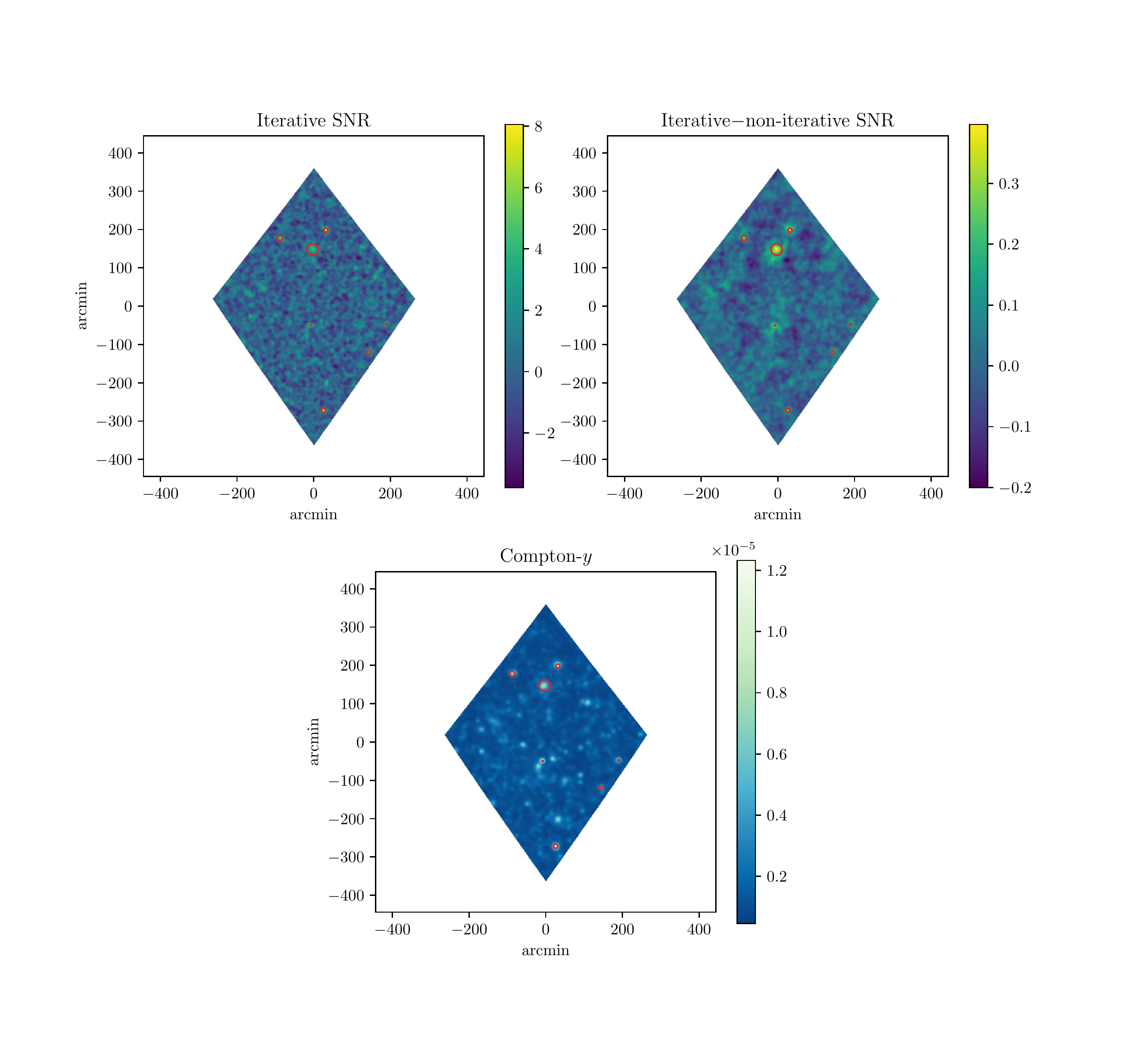}
\\
\caption{\textit{Upper-left panel}: Iterative signal-to-noise map at $\theta_{500} = 7.24$\,arcmin of a reference tile for our \textit{Planck}-like experiment. Detections are shown as red circles. \textit{Upper-right panel}: Difference between the iterative signal-to-noise map and its non-iterative counterpart for the same tile. \textit{Lower panel}: Input Compton-$y$ map, from the Websky simulation \citep{Stein2020}.} 
\label{fig:snrmap}
\end{figure*}

Our detection pipeline is illustrated in Figure~\ref{fig:snrmap} for a reference tile of a \textit{Planck}-like experiment (see Section \ref{sec:sims} for the specification details). The upper-left panel shows the iterative signal-to-noise map at $\theta_{500} = 7.24$\,arcmin, with detections shown as red circles, the circle radii being the estimated values of $\theta_{500}$. The upper-right panel shows, for the same angular scale, the difference between the iterative signal-to-noise map and its non-iterative counterpart (the non-iterative map itself is not shown due to it being visually very similar to the iterative one). A signal-to-noise gain is observed around the detections, as expected from Section \ref{subsec:ps}. Finally, the lower panel shows the input Compton-$y$ map (from the Websky simulation, \citealt{Stein2020}). It is clear that detections correspond to real halos.

\subsection{Merging of tile catalogues}
The final step of the iMMF cluster detection method involves merging the detections from the different tiles into a single catalogue of cluster candidates (iterative or non-iterative) and, finally, merging detections that are within some distance of each other. The first step is trivial and leads to a single survey-wide catalogue, in which detections are assigned global sky coordinates (Galactic longitude and latitude), as well as a number identifying the tile in which they were found. In the second step, a friends-of-friends algorithm is run on this survey-wide catalogue, identifying detections that are within a given angular distance $\theta_{\mathrm{id}}$ of any other detection as the same object. For each friends-of-friends cluster, the values of $q_{\mathrm{opt}}$, $\hat{y}_0$, $\theta_{500}$ and sky location of the member with the highest value of $q_{\mathrm{opt}}$ are assigned to the merged detection, which is the one that goes into the final catalogue. For our analysis of simulated \textit{Planck}-like data, we choose $\theta_{\mathrm{id}} = 10$\,arcmin. 
Merging detections this way solves two problems. First, our peak-finding algorithm, described in Section \ref{subsec:peaks}, sometimes leads to several detections associated to a single true object, as it finds local peaks in the signal-to-noise three-dimensional distribution, which is noisy. These are always, however, very close to each other, and the friends-of-friends algorithm successfully merges them into a single detection. In addition, clusters lying very close to the border between two tiles may be detected in both tiles (or in more than two, if the cluster is near the intersection between four tiles). The friends-of-friends algorithm also successfully merges these detections into a single one.

\section{Mock observations}\label{sec:sims}
In this section, we describe the mock data on which we apply our iMMF. We first describe the simulated maps that we use and the \textit{Planck}-like experimental specifications that we assume. Next, we detail the specific cluster catalogues that we produce with our iMMF implementation, which are then analysed in Section \ref{sec:results}.

\subsection{Sky maps}
We consider one single full sky, observed at the six highest frequency channels of \textit{Planck} and in the HEALPix pixelisation with $N_{\mathrm{side}} = 2048$, which is the $N_{\mathrm{side}}$ used in the legacy \textit{Planck} maps. Each of our six all-sky temperature maps contain the following components, all of which are convolved with the corresponding \textit{Planck} isotropic beam and the HEALPix pixel transfer function  (except the instrumental noise, which is only convolved with the pixel transfer function):
\begin{itemize}
    \item \textbf{tSZ:} We use two different Compton-$y$ maps. The first one is that from the publicly-available Websky simulation\footnote{\texttt{mocks.cita.utoronto.ca/websky}} \citep{Stein2019,Stein2020}, an all-sky second-order Lagrangian perturbation theory lightcone simulation with a minimum halo mass $M_{200} \sim 1.4 \times 10^{12} M_{\odot}$. Since the Websky Compton-$y$ map has $N_{\mathrm{side}} = 4096$, we degrade it to $N_{\mathrm{side}} = 2048$ with the HEALPix \texttt{ud\_grade} function. We degrade in a similar way all the Websky maps that we use (kSZ, lensed CMB and CIB). In addition, we produce a Compton-$y$ map with a custom-generated cluster catalogue, which is described below. For this map, we assume the same cluster profile that is used in our MMF implementation, i.e., that of \citet{Arnaud2010} (see Section \ref{sec:mmf}), and paint the clusters at uncorrelated random locations. We note that we will only use this `injected' map in combination with the Websky one.
    \item \textbf{kSZ:} We use the kinetic SZ (kSZ) signal of the Websky simulation, which includes both the late-time and the reionisation constributions \citep{Stein2020}.
    \item \textbf{Lensed CMB:} Also taken from the Websky simulation.
    \item \textbf{CIB:} The Cosmic Infrared Background (CIB) signal is similarly taken from the Websky simulation. The CIB is expected to be spatially correlated with the tSZ signal, which may have a non-negligible impact on the MMF extraction. Since this work focuses on the impact of noise covariance misestimation, we decide to spatially decorrelate the CIB with respect to the tSZ signal, leaving the study of the tSZ-CIB correlation to further work. In particular, when analysing each individual tile, we assign a random CIB cut-out to it, instead of the corresponding one.
    \item \textbf{Instrumental noise:} We generate a map of Gaussian white noise for each channel, with noise levels, for increasing frequencies, of 77.4\,$\mu$K\,arcmin, 33\,$\mu$K\,arcmin, 46.8\,$\mu$K\,arcmin, 153.6\,$\mu$K\,arcmin, 46.8\,kJy\,sr$^{-1}$\,arcmin, and 43.2\,kJy\,sr$^{-1}$\,arcmin \citep{Planck2016viii}. The noise is uncorrelated between different channels, and, as noted above, is convolved only by the HEALPix pixel transfer function, and not by the instrument beam.  
\end{itemize}
We note that our maps do not include Galactic foregrounds and radio point sources. As a consequence, in our analysis we do not impose neither a Galactic mask nor a point source mask, i.e., we use the full sky.

\begin{figure}
\centering
\includegraphics[width=0.5\textwidth]{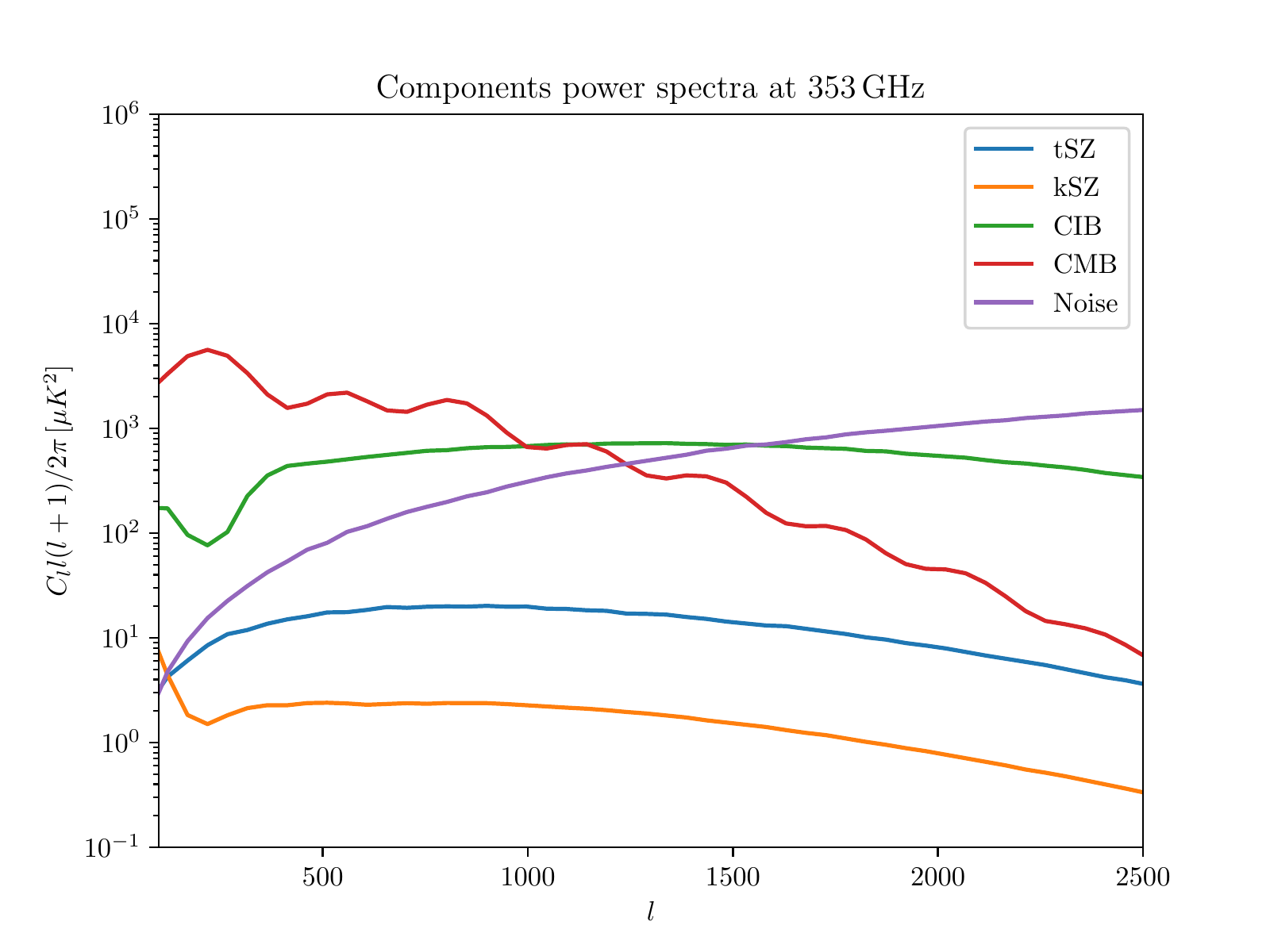}
\caption{Power spectra of the different components in our 353\,GHz map. The tSZ component is that from the Websky simulation.}
\label{fig:powerspectrum}
\end{figure}

Figure \ref{fig:powerspectrum} shows the power spectra of the different components in our 353\,GHz map for our Websky maps, computed from 100 of the square cut-outs described in Section \ref{subsec:tesselation}. Note that the tSZ power spectrum is small compared to CMB, CIB and noise contributions. However, as the MMF is designed to look for the tSZ signal, spatially but also spectrally, in Section \ref{sec:results} we will see that it has a non-negligible impact on MMF noise covariance estimation.

\subsubsection{Custom $y$ map}
In order to produce our custom Compton-$y$ map, we generate a cluster catalogue by sampling the halo distribution in a mass range between $M_{500} = 4 \times 10^{14} M_{\odot}$ and $M_{500} = 2 \times 10^{15} M_{\odot}$ and a redshift range between $z=0$ and $z=2$. We draw a total of 4000 samples. Whilst this is not a completely realistic catalogue, the motivation is to have a tSZ map that only contains massive clusters with a relatively high chance of being detected. A typical tile will contain $4000/768 = 5.2$ clusters. This is in contrast with the Websky tSZ map, which has halos down to masses well below the \textit{Planck} detection threshold, with a typical tile containing hundreds of halos. 



\subsection{Cluster catalogues}
We apply our iMMF implementation, described in Section \ref{sec:mmf}, to our simulated all-sky \textit{Planck}-like maps, producing the following catalogues:
\begin{itemize}

\item An iterative and a non-iterative catalogue with the Websky tSZ signal, with the MMF in the cluster-finding mode (`iterative Websky' and `non-iterative Websky' catalogues).
\item A non-iterative catalogue with the Websky tSZ signal and the tSZ maps not present in the multifrequency maps used to estimate the noise covariance (`true noise Websky'). We will use this catalogue in order to investigate the contribution from the undetected tSZ signal to the MMF variance.
\item An iterative and non-iterative catalogues using the Websky tSZ signal in addition to our custom injected tSZ signal, with the MMF in the cluster-finding mode, keeping only the detections corresponding to the injected clusters (`iterative Websky+injected' and `non-iterative Websky+injected' catalogues).
\item An iterative and non-iterative catalogues with our custom injected tSZ signal in addition to our custom injected tSZ signal, with the MMF in fixed mode, extracting the signal-to-noise at the true cluster parameters of all the injected clusters (`fixed iterative Websky+injected' and `fixed non-iterative Websky+injected' catalogues). This catalogue also includes the true signal-to-noise of the detections, computed by applying our MMF to the injected tSZ map at the true cluster parameters.

\end{itemize}
Before analysing the properties of the catalogues, we impose a new signal-to-noise threshold of 5 in all of them except in our fixed Websky+injected catalogues, for which we keep all the injected clusters.

\section{Results and discussion}\label{sec:results}
In this section, we use our \textit{Planck}-like cluster catalogues in order to investigate the impact of the presence of the cluster tSZ signal in the MMF noise covariance on MMF extraction and the potential of iterative covariance estimation to mitigate against it. First, we quantify the gain in signal-to-noise provided by our iterative approach. We then look at the biases in the mean and the standard deviation of the signal-to-noise observable when the covariance is misestimated, and at how iterative estimation can suppress them. Finally, we quantify the completeness and the purity of our catalogues, both of them essential ingredients in a cluster number count likelihood.

\subsection{Signal-to-noise gain}\label{subsec:gain}
\begin{figure}
\centering
\includegraphics[width=0.5\textwidth]{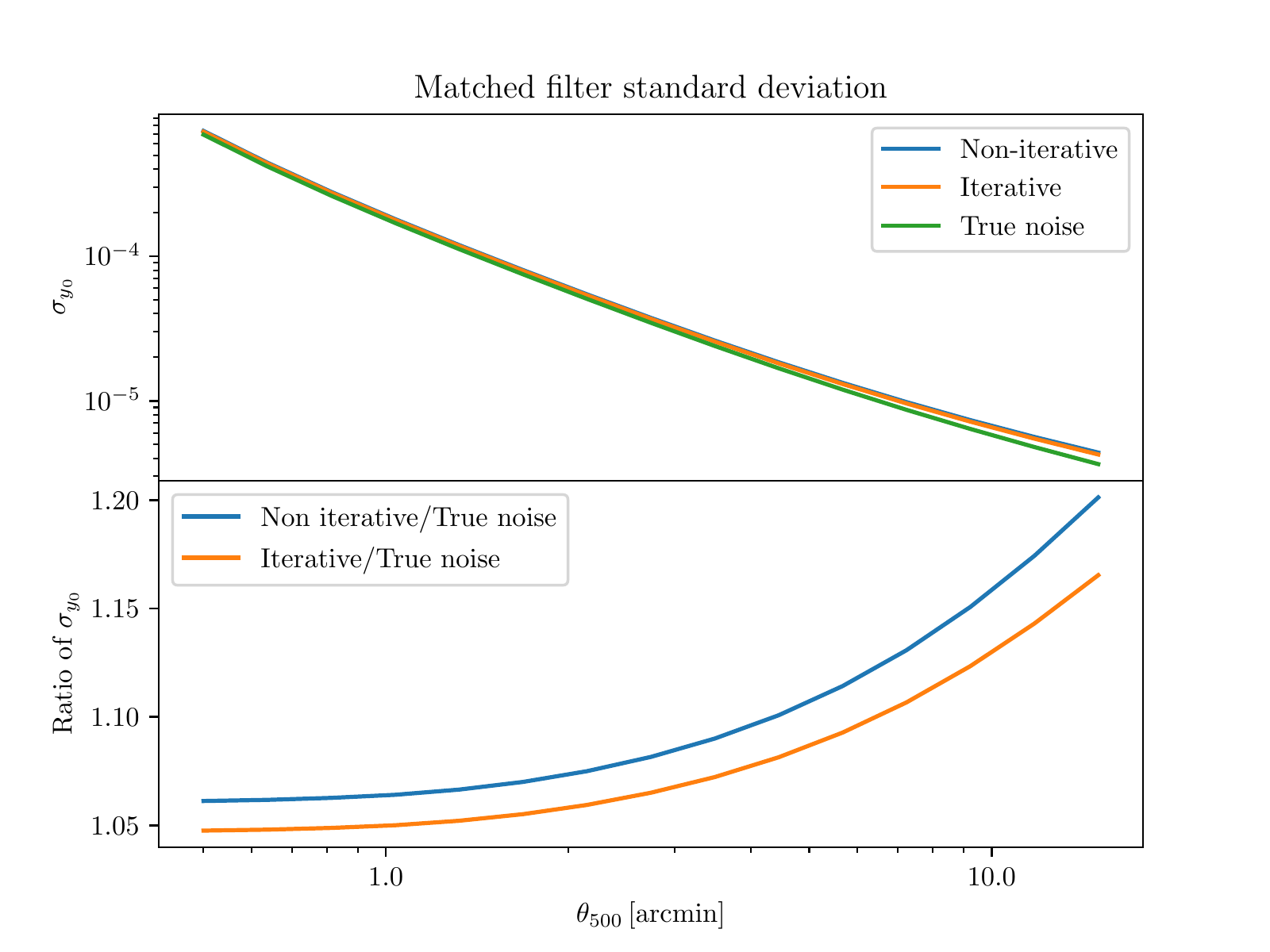}
\caption{Matched filter standard deviation as a function of the MMF angular scale $\theta_{500}$ for the `true noise', iterative, and non-iterative Websky MMFs, averaged over 100 tiles.}
\label{fig:mf_std}
\end{figure}

The upper panel of Figure~\ref{fig:mf_std} shows the matched filter standard deviation as a function of filter angular scale, $\sigma_{y_0} (\theta_{500})$, for our three Websky MMFs: the `true noise', the iterative, and the non-iterative MMFs. The lower panel, on the other hand, shows, for the two latter MMFs, the ratio of $\sigma_{y_0} (\theta_{500})$ to that of the true noise MMF. We recall that $\sigma_{y_0}$ is given $N^{-1/2}$, where $N$ is the matched filter normalisation (see Section~\ref{sec:mmf}). For a given central Compton-$y$ parameter, $\sigma_{y_0}$ sets the detection signal-to-noise, $q = y_0/\sigma_{y_0}$. We note that the curves in Figure~\ref{fig:mf_std} are computed by averaging over 100 tiles. 

As expected, the non-iterative MMF has the largest associated standard deviation, as the full tSZ signal contributes to it, followed by the iterative MMF, in which the brightest clusters are removed for covariance estimation. Finally, the true noise MMF yields the smallest standard deviation, as the tSZ signal does not contribute to it at all. The difference between the non-iterative and the iterative MMF noise curves is thus due to the contribution of the brightest, most massive clusters to the MMF standard deviation. On the other hand, the difference between the iterative and the true MMF noise curves is due to the contribution of the sea of less bright, undetected halos. Figure \ref{fig:mf_std} clearly shows that, although they may not be detected individually, for a \textit{Planck}-like experiment they contribute significantly to the MMF standard deviation.

\begin{figure}
\centering
\includegraphics[width=0.5\textwidth]{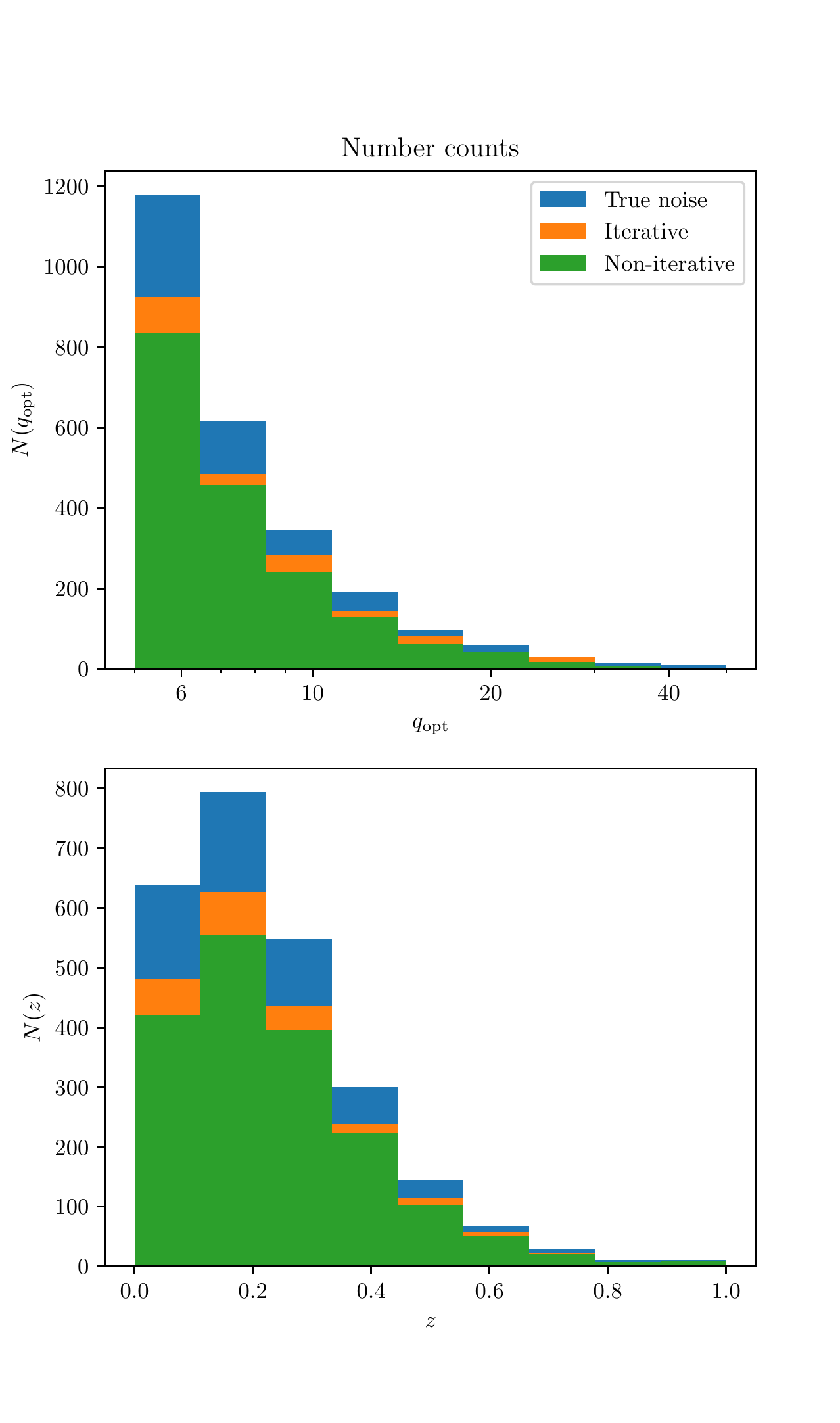}
\caption{Number counts of the detections as a function of detection, or optimal, signal-to-noise $q_{\mathrm{opt}}$ (upper panel) and redshift (lower panel) for the true noise, iterative, and non-iterative Websky catalogues.}
\label{fig:histogram}
\end{figure}

Figure \ref{fig:histogram} shows how these MMF standard deviation curves translate into detected number counts. Here, the number counts of the three Websky catalogues are shown as a function of the detection, or optimal, signal-to-noise $q_{\mathrm{opt}}$ (upper panel), as well as a function of redshift (lower panel). As expected from Figure \ref{fig:mf_std}, the non-iterative MMF leads to the lowest number of detections, followed by the iterative MMF and the true noise MMF. It is clear that iterative covariance estimation leads to more detected clusters, and hence to an enhanced statistical power, relative to the non-iterative case.

We note that the true noise MMF, which leads to the lowest standard deviation and, correspondingly, to the highest number of detections, should be thought of as an idealised limit, useful only to quantify the `noise floor' due to the sea of undetected halos. Even if the contribution of these halos to the covariance could be determined from the data and removed, leading to a significantly higher number of detections, the variability due to them would still be present in the data. Hence, the MMF estimator ($\hat{y}_0$ or $q_{\mathrm{opt}}$) would still be affected by them. Indeed, although what is understood as noise in the MMF would decrease relative to the non-iterative and the iterative cases, the intrinsic scatter of the observable would necessarily increase, leading to no real improvement in the statistical power of the data. This is alike to, e.g., only thinking of instrumental noise as noise in the MMF: it can be done (determining the noise covariance from, e.g., half-mission maps), and it would lead to more detections, but there would be no real signal-to-noise gain, as the data would still know about all the other components (the CMB, the CIB, etc.). We remark that masking the detections, however, as is done in our iterative approach, is different: in this case, it is what is rightly understood as the signal that is being removed from the covariance estimate.

\begin{figure}
\centering
\includegraphics[width=0.5\textwidth]{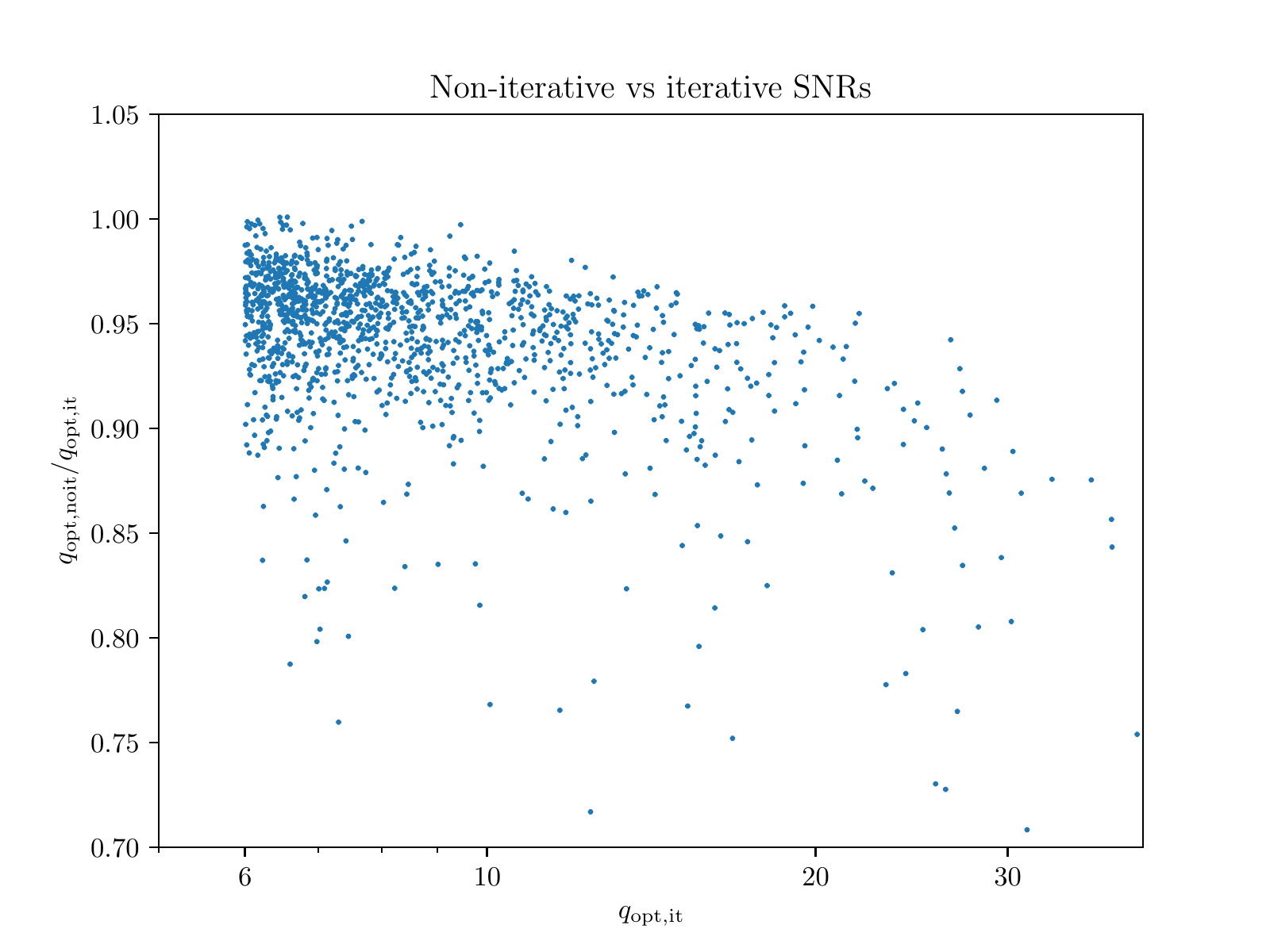}
\caption{Ratio of the non-iterative signal-to-noise measurements to their iterative counterparts for the Websky clusters. The signal-to-noise gain due to iterative covariance estimation is evident.}
\label{fig:ratio_snrs}
\end{figure}

We can also look at the signal-to-noise change of the individual detections in the iterative catalogue relative to their corresponding non-iterative values. Figure \ref{fig:ratio_snrs} shows the ratio of the non-iterative $q_{\mathrm{opt}}$ measurements to their non-iterative counterparts plotted against the latter. The signal-to-noise gain is evident, increasing from a few percent at low signal-to-noise to $\simeq 15$\,\% or more for the most significant detections. By adding all the signal-to-noise measurements of each catalogue in quadrature, we report an overall signal-to-noise gain of 9.1\,\% for the iMMF relative to the non-iterative MMF.

\subsection{Bias on mean and standard deviation}\label{subsec:resultsbias}
We now turn to our injected+Websky cluster catalogues in order to investigate the mean and standard deviation of the main cluster observable, the signal-to-noise. We will also use these catalogues to quantify the sample completeness in Section \ref{subsec:completeness}. We use our injected+Websky catalogues because we need to measure the true signal-to-noise of each cluster, and for them, we have access to the individual injected tSZ signals. This is not the case for the Websky clusters, as the Websky Compton-$y$ map has contributions from all the undetected clusters, which are thought of as noise in our MMFs. We recall that the injected+Websky catalogues only include the measurements corresponding to the injected clusters, and not those associated to Websky's (note, however, that the input tSZ map has both contributions).

\begin{figure*}
\centering
\includegraphics[width=0.8\textwidth]{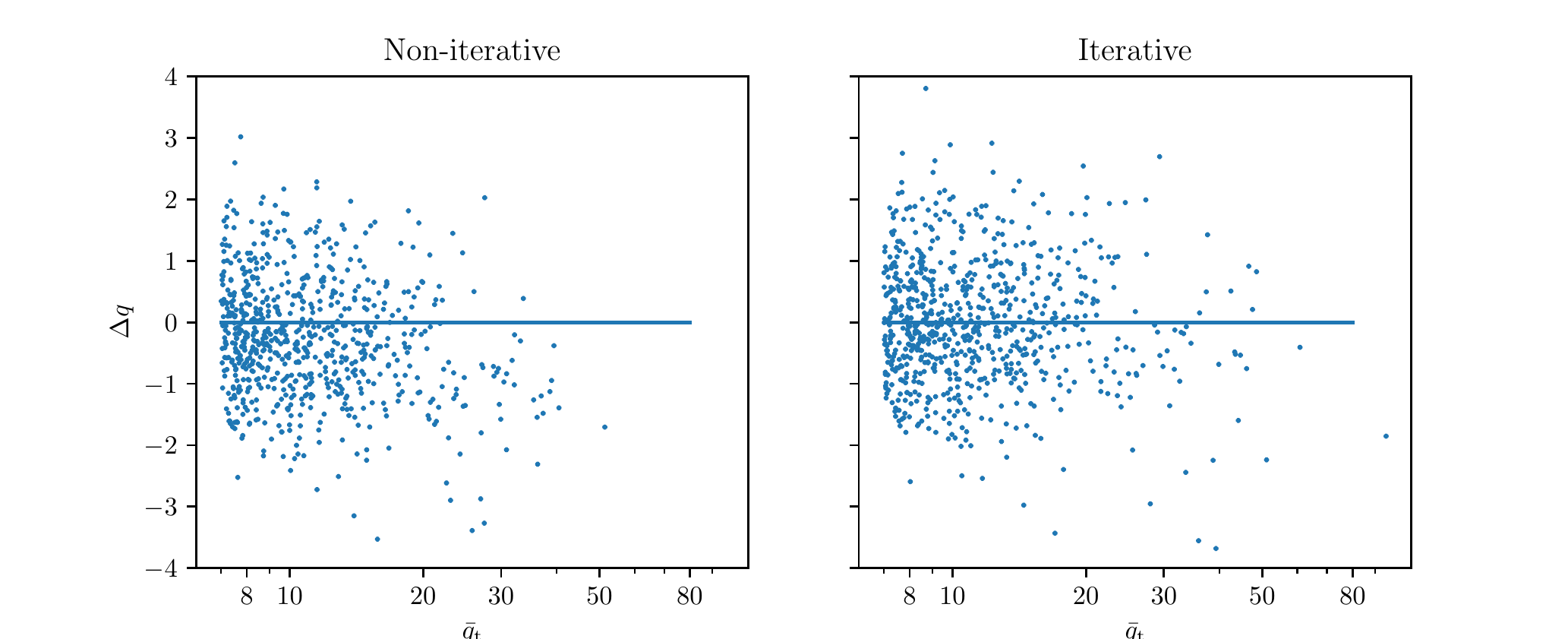}
\caption{Signal-to-noise residuals of the injected clusters in the injected+Websky catalogues (non-iterative on the left, and iterative on the right). A negative bias is clearly visible in the non-iterative case, particularly at high $\bar{q}_{\mathrm{t}}$, whereas no bias is visually apparent in the iterative case.}
\label{fig:scatter}
\end{figure*}

Figure \ref{fig:scatter} shows the signal-to-noise residuals $\Delta q$ as a function of the true signal-to-noise $\bar{q}_{\mathrm{t}}$ for our non-iterative injected+Websky catalogue (left panel) and our iterative injected+Websky catalogue (right panel). The residuals are defined as $\Delta q = q_{\mathrm{opt}} -  (\bar{q}_{\mathrm{t}}^2 + 3)^{1/2}$ for the optimal signal-to-noise $q_{\mathrm{opt}}$, i.e., if the clusters are blindly detected (which is the case shown in Figure \ref{fig:scatter}), and as $\Delta q = q - \bar{q}_{\mathrm{t}}$ if the signal-to-noise is measured at the true cluster parameters ($\theta_{500}$ and sky location). The former definition accounts for the optimisation bias, as discussed extensively in \citet{Zubeldia2021}. According to our theoretical understanding of matched filters (see Sections \ref{sec:mmf} and \ref{subsec:ps}), both sets of residuals should have zero mean and unity standard deviation if the noise covariance is correctly determined. In the left panel of Figure \ref{fig:scatter}, a negative bias is clearly observed. This bias is not visually apparent in right panel, which corresponds to the iterative catalogue.

\begin{figure*}
\centering
\includegraphics[width=0.8\textwidth]{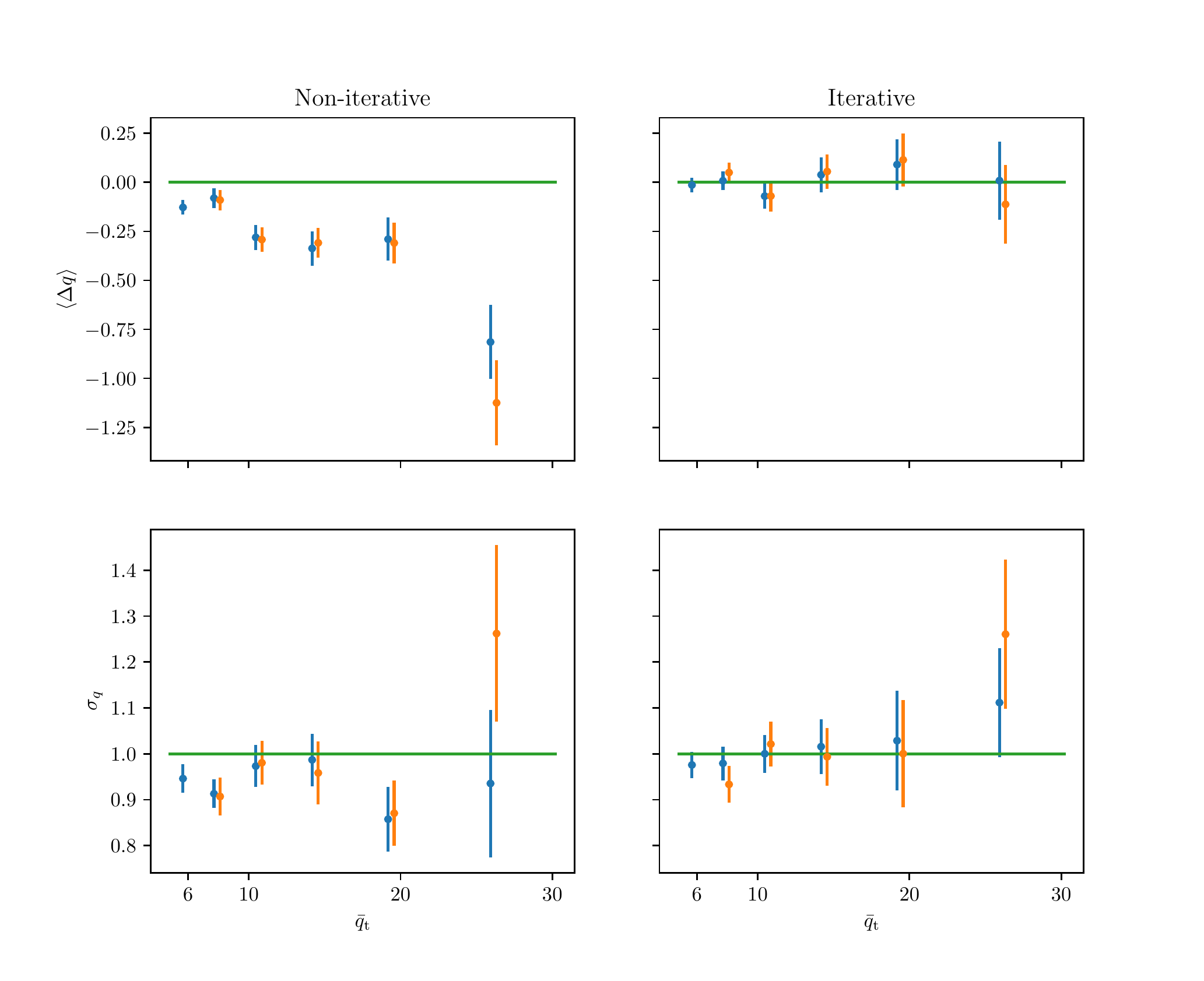}
\caption{Empirical mean and standard deviation (upper and lower panels, respectively) of the signal-to-noise measurements of our injected+Websky catalogues binned in true signal-to-noise for both non-iterative (left panels) and iterative (right panel) covariance estimation. The blue points correspond to the `fixed' case, in which the signal-to-noise is extracted at the true cluster parameters, whereas the orange points correspond to the optimal signal-to-noise, obtained blindly by maximisation over angular scale and sky location. A bias in both $\Delta q$ and $\sigma_q$ is observed in the non-iterative case, but they are compatible with no bias and unit value, respectively, in the iterative case.}
\label{fig:mean_std}
\end{figure*}

More quantitatively, Figure \ref{fig:mean_std} shows the mean and standard deviation of the residuals (upper and lower panels, respectively), binned in true signal-to-noise, for the non-iterative and iterative catalogues (left and right panels, respectively), both if the clusters are blindly detected (orange points) and if the signal-to-noise is extracted at the true cluster parameters (blue points). The error bars are obtained by bootstrapping. A negative bias is observed in the non-iterative catalogues, increasing in magnitude with signal-to-noise. No bias is seen in the iterative catalogues. For the non-iterative catalogues, the mean bias across all the bins is $\left\langle \Delta q \right\rangle = -0.270 \pm 0.036$ and $\left\langle \Delta q \right\rangle = -0.193 \pm 0.026$ for the cluster finding and fixed catalogues, respectively, which are -7.5\,$\sigma$ and -7.2\,$\sigma$ away from zero, respectively. This can be seen as the significance by which the overall, or stacked, cluster observable, the signal-to-noise, is away from its predicted value. Thus, our non-iterative catalogues could potentially lead to biased cosmological constraints unless this bias is accounted for. For the iterative catalogues, on the other hand, we report a mean bias of $\left\langle \Delta q \right\rangle = -0.011 \pm 0.036$ and $\left\langle \Delta q \right\rangle = -0.008 \pm 0.026$ for the cluster finding and fixed catalogues, respectively, consistent with zero in both cases. In a cosmological analysis, we would thus expect the iMMF to yield unbiased constraints (provided that all the other potential systematics in the analysis are properly accounted for).

We note that the errors on the mean bias for the cluster-finding catalogues are larger than their fixed counterparts because the first signal-to-noise bin, $5 < \bar{q}_{\mathrm{t}} < 7$, is omitted in the cluster-finding case. Similarly, in the first signal-to-noise bins of Figure \ref{fig:mean_std} only the fixed catalogue values are shown. This is because the optimal signal-to-noise measurements suffer from a significant Malmquist bias in this bin, as the cluster-finding catalogues are selected by imposing a signal-to-noise threshold at $q_{\mathrm{opt}} = 5$: at fixed $\bar{q}_{\mathrm{t}}$, clusters with a preferentially high value of $q_{\mathrm{opt}}$ are detected. This effect is, however, not present in the fixed catalogues, as for them the signal-to-noise is measured for all the clusters in the catalogue, regardless of whether they fall below the selection threshold or not. Note that the Malmquist bias is negligible above $\bar{q}_{\mathrm{t}} \simeq 7$, as in this regime most clusters are detected. We remark that this Malmquist bias would be of no concern in a likelihood analysis, in which it would be naturally accounted for if the sample selection is properly modelled via the completeness function (see Section \ref{subsec:completeness}).

The bias that we see in our non-iterative catalogues is the covariance bias that we predicted in Section \ref{subsubsec:noise} and observed with a toy model in Section \ref{sec:toymodel}. It is caused by the contribution of the detected clusters to the noise covariance estimate. We note that, for each catalogue (iterative and non-iterative), the true signal-to-noise is computed with the same covariance that is used to compute the observed signal-to-noise. This is consistent with what would be done in a likelihood analysis. Thus, the covariance bias is unrelated to the fact that the non-iterative signal-to-noise measurements are systematically lower to their iterative counterparts, as reported in Section \ref{subsec:gain} (see, e.g., Figure \ref{fig:ratio_snrs}).


As well as being biased, Figure \ref{fig:mean_std} also shows that the non-iterative signal-to-noise has a standard deviation that is consistently less than unity. This is due to the noise covariance being overestimated, as predicted in Section \ref{subsubsec:noise}. Assuming it to be unity in a likelihood analysis would therefore constitute an error in the modelling of the observable. This could potentially lead to a bias in the constraints, and would certainly lead to a loss of constraining power, as the observable would be assumed to be more noisy than it actually is, as explained in Section \ref{subsec:ps}. For the iterative catalogues, on the other hand, the signal-to-noise standard deviation is observed to be consistent with unity throughout. 

In summary, the presence of the cluster signal in the covariance estimate leads to signal-to-noise measurements that are biased low with respect to their predicted values and whose standard deviation is less than unity, the value that would be naively predicted. Our iterative approach is able to suppress both effects completely, leading to a signal-to-noise observable that behaves as expected. Namely, it is unbiased with respect to the prediction, i.e., the true signal-to-noise, and has unity standard deviation. Our iterative approach thus eliminates any potential biases in the cosmological inference caused by these effects. As noted in Section \ref{subsec:ps}, we remark that this is true regardless of whether the model, i.e., the MMF template, is a perfect fit to the data. Indeed, any modelling errors are absorbed into the definition of the true signal-to-noise. Iterative noise estimation does not address this issue, which remains to be taken care of if unbiased constraints are to be obtained.

Finally, we highlight the overall excellent agreement between the residuals of the optimal and the fixed catalogues, with the fluctuations in the mean and standard deviation being highly correlated. This fact, along with the excellent agreement of the optimal signal-to-noise residuals with having zero mean and unity standard deviation, further validates the modelling of the optimal signal-to-noise as unit-variance Gaussian centred at $(\bar{q}_{\mathrm{t}}^2 + 3)^{1/2}$, as argued in \citet{Vanderlinde2010} and, in more detail, in \citet{Zubeldia2021}. 


\subsection{Completeness}\label{subsec:completeness}
\begin{figure*}
\centering
\includegraphics[width=0.8\textwidth]{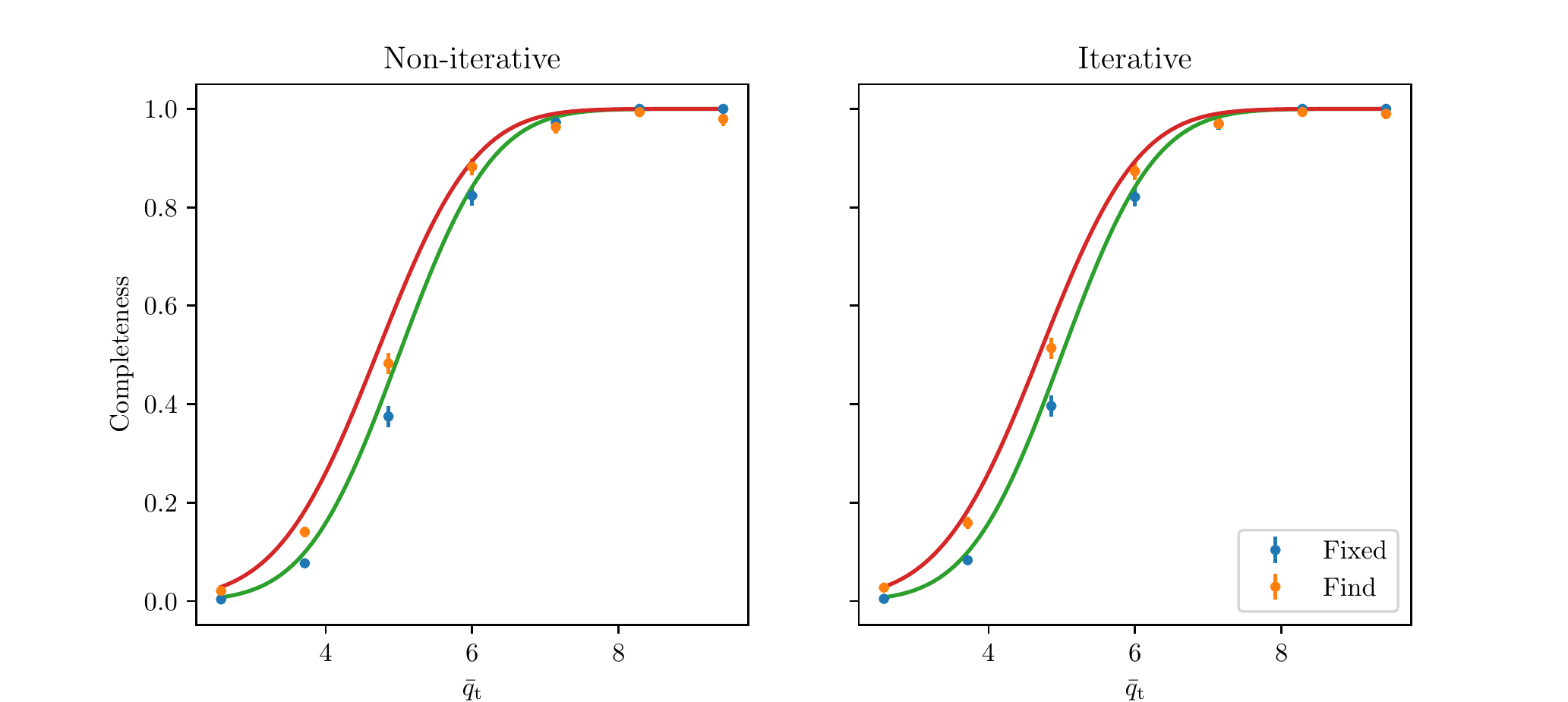}
\caption{Completeness of our injected+Websky catalogues, both for non-iterative and iterative noise estimation (left and right panels, respectively), and for blind cluster finding and extraction at the true cluster parameters (orange and blue points, respectively). The predictions from Eq. \ref{erf} are also shown (red curve for the cluster finding case, and green curve for the fixed case).}
\label{fig:completeness}
\end{figure*}

Let us now consider the completeness of our injected+Websky cluster catalogues. The completeness of a catalogue is defined as the probability for a cluster to be included in the catalogue at a given value of the true signal-to-noise, $P(\mathrm{i}|\bar{q}_{\mathrm{t}})$, where i denotes the inclusion of the cluster in the catalogue. Consider two catalogues, one with optimal signal-to-noise measurements $q_{\mathrm{opt}}$, produced by a blind cluster search, and one with fixed signal-to-noise measurements extracted at the true cluster parameters, $q_{\mathrm{fix}}$. Both catalogues are constructed by imposing a signal-to-noise threshold $q_{\mathrm{th}}$, i.e., $P(\mathrm{i}|q) = \Theta (q - q_{\mathrm{th}} )$, where $q$ is either $q_{\mathrm{opt}}$ or $q_{\mathrm{fix}}$, and $\Theta(x)$ is the step function. Assuming each measurement follows a unit-variance Gaussian centred at $\bar{q}$, where $\bar{q} = \bar{q}_{\mathrm{t}}$ in the fixed case and $\bar{q} = (\bar{q}_{\mathrm{t}}^2 + 3)^{1/2}$ in the optimal case, the completeness is then given by (e.g., \citealt{Ade2016})
\begin{equation}\label{erf}
P(\mathrm{i}|\bar{q}_{\mathrm{t}}) = \frac{1}{2} \left[ 1 - \mathrm{erf} \left( \frac{q_{\mathrm{th}} - \bar{q}}{\sqrt{2}} \right) \right].
\end{equation}
Figure~\ref{fig:completeness} shows the empirical completeness of the injected clusters in our injected+Websky catalogues, both with non-iterative and iterative covariance estimation (left and right panels, respectively), and with blind cluster finding (`find' case, orange points) and extraction at the true cluster parameters (`fixed' case, blue points). In addition, the theoretical prediction of Eq. (\ref{erf}) is shown for both the find and the fixed cases (red and green curves, respectively). We recall that the blind cluster catalogues are constructed by setting the signal-to-noise threshold at $q_{\mathrm{th}}=5$. Here, we also impose a threshold of $q_{\mathrm{th}}=5$ for our fixed catalogues, which, we recall, originally contain the signal-to-noise measurements of all the injected clusters. 

Very good agreement between the theory and the data is found for the iterative case, with perhaps a small bias at around $\bar{q}_{\mathrm{t}} \sim 5$. In the non-iterative case, however, the data points are significantly below the prediction for both the fixed and cluster-finding catalogues. This can be easily understood as due to the covariance bias: at given $\bar{q}_{\mathrm{t}}$, clusters have an observed signal-to-noise that is biased low with respect to $\bar{q}$ (see Figure \ref{fig:mean_std}). Thus, they have a lower chance of being detected with respect to what is predicted if the bias is ignored. Similarly to what we found in our analysis of the signal-to-noise mean and standard deviation, this bias in the completeness is suppressed with iterative covariance estimation.

Apart from showing the power of the iterative approach in reducing the bias in the completeness, Figure \ref{fig:completeness} also provides further evidence for the existence of the optimisation bias that was extensively discussed in \citet{Zubeldia2021}, as well as for it being well approximated by the simple correction $\bar{q} = (\bar{q}_{\mathrm{t}}^2 + 3)^{1/2}$.

\subsection{Purity}\label{subsec:purity}

\begin{figure}
\centering
\includegraphics[width=0.5\textwidth]{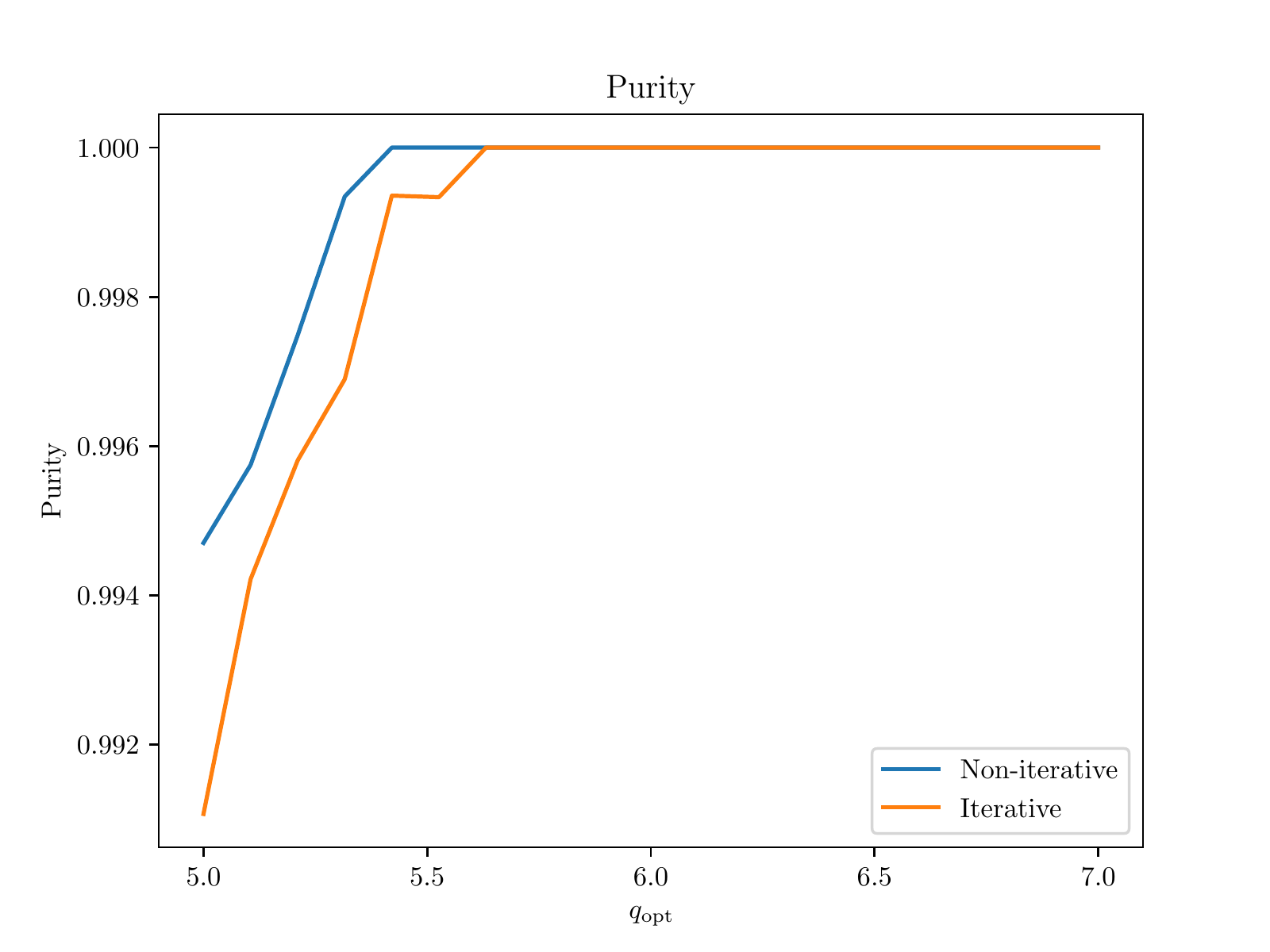}
\caption{Purity, defined as the number of false detections over the total number of detections (purity) in our Websky catalogues, both for non-iterative and iterative noise covariance estimation (blue and orange curves, respectively). There is a small reduction in the purity in going from non-iterative to iterative.}
\label{fig:purity}
\end{figure}

Finally, we investigate the purity, or reliability, of our cluster catalogues., i.e., the probability of there being false detections in them. We do so with our Websky catalogues, cross-matching our detections with the input Websky halo catalogue. Figure \ref{fig:purity} shows the number of false detections over the total number of detections as a function of minimum detection, or optimal, signal-to-noise $q_{\mathrm{opt}}$, for both the non-iterative and the iterative Websky catalogues. Both catalogues display an excellent purity, being more than 99\,\% pure above $q_{\mathrm{opt}}=5$ and 100\,\% pure above $q_{\mathrm{opt}}=6$. The purity of the iterative catalogue is slightly lower than that of the non-iterative catalogue, which is not surprising, as iterative noise estimation lowers the MMF noise, which can potentially lead to more false detections. It is, however, still excellent.

\section{Conclusion}\label{sec:conclusion}

Multi-frequency matched filters (MMFs) are routinely used in order to detect blindly galaxy clusters from CMB data. They have indeed been applied to \textit{Planck}, ACT, and SPT data (e.g., \citealt{Bleem2015,Planck2016xxvii,Hilton2020}), and will also be used in SO and CMB-S4. Given the great statistical power of these upcoming experiments, which are expected to deliver over an order of magnitude more clusters than their predecessors, systematics in the cluster detection methods will have to be understood more accurately than ever. This work, in which we have investigated the impact of covariance contamination by the cluster signal on cluster MMF detection, constitutes a step in this direction. 

MMFs require an estimate for the map noise covariance in order to be applied. Thus far, the noise covariance is typically estimated from the data and taken to be equal to the data covariance (e.g., \citealt{Ade2016}). In this work, we have shown that the presence of the detected clusters' tSZ signal in the covariance estimate has a non-negligible impact on MMF cluster detection. Specifically, we have shown that (1) it leads to a signal-to-noise loss, as the noise covariance is overestimated, degrading the cosmological inference, and (2) it leads to a cluster observable (the signal amplitude $\hat{y}_0$ or the signal-to-noise $q$) which does not behave as predicted, potentially leading to biased cosmological inference. In particular, the observable is biased low with respect to its theoretically-predicted value, a bias which we have called `covariance bias', and it has a standard deviation that is smaller than its theoretical prediction (in the case of the signal-to-noise, less than unity). We have provided theoretical arguments for the existence of these effects (Section \ref{subsec:ps}) and we have quantified them numerically, first with a toy model (Section \ref{sec:toymodel}), and then with realistic \textit{Planck}-like cluster catalogues (Section \ref{sec:results}).

We have proposed a novel approach, the iterative MMF (iMMF), designed to suppress these effects. In our iMMF, a cluster catalogue is first constructed by taking the noise covariance to be equal to the data covariance, as customary. Then, the noise covariance is reestimated by masking from the data the detections above some signal-to-noise threshold. This procedure, which can be iterated a number of times, leads to a new, iterative catalogue, in what constitutes a clean and model-agnostic way of removing the contamination of the noise covariance by the detections. Masking, as opposed to subtraction of a model, was found to be the preferred procedure, since subtraction is intrinsically inaccurate, leading to residuals in the maps, and potentially numerically unstable.

We have then implemented our iMMF method numerically and applied it to realistic \textit{Planck}-like data, producing a number of mock cluster catalogues. With them, we have shown that, after just one iteration, our iMMF boosts the detection signal-to-noise, leading to more and more significant detections (Section \ref{subsec:gain}), with an overall signal-to-noise gain of 9.1\,\%. We have also shown that the iMMF eliminates completely the covariance bias and the deviation of the signal-to-noise standard deviation from unity (Section \ref{subsec:resultsbias}), leading to a cluster observable that behaves as predicted by the theory, and thus eliminating potential biases in the cosmological inference. In addition, as a consequence of this, our iMMF also brings the catalogue completeness function in agreement with its theoretical prediction (Section \ref{subsec:completeness}). We have also verified that it has a very small impact on the number of false detections (Section \ref{subsec:purity}). Given the implementational simplicity of our extension to the MMF cluster finding algorithm and its clear benefits, we recommend for it to be used in any future applications of the MMF algorithm to real data.


\section*{Acknowledgements}
This work was supported by the ERC Consolidator Grant {\it CMBSPEC} (No.~725456) as part of the European Union's Horizon 2020 research and innovation program. IZ was also supported by an STFC Consolidated Grant, and JC by the Royal Society as a Royal Society URF at the University of Manchester

\section*{Data Availability}

The data underlying this article will be shared on reasonable request to the corresponding author.



\bibliographystyle{mnras}
\bibliography{references} 

\begin{thebibliography}{}
\makeatletter
\relax
\def\mn@urlcharsother{\let\do\@makeother \do\$\do\&\do\#\do\^\do\_\do\%\do\~}
\def\mn@doi{\begingroup\mn@urlcharsother \@ifnextchar [ {\mn@doi@}
  {\mn@doi@[]}}
\def\mn@doi@[#1]#2{\def\@tempa{#1}\ifx\@tempa\@empty \href
  {http://dx.doi.org/#2} {doi:#2}\else \href {http://dx.doi.org/#2} {#1}\fi
  \endgroup}
\def\mn@eprint#1#2{\mn@eprint@#1:#2::\@nil}
\def\mn@eprint@arXiv#1{\href {http://arxiv.org/abs/#1} {{\tt arXiv:#1}}}
\def\mn@eprint@dblp#1{\href {http://dblp.uni-trier.de/rec/bibtex/#1.xml}
  {dblp:#1}}
\def\mn@eprint@#1:#2:#3:#4\@nil{\def\@tempa {#1}\def\@tempb {#2}\def\@tempc
  {#3}\ifx \@tempc \@empty \let \@tempc \@tempb \let \@tempb \@tempa \fi \ifx
  \@tempb \@empty \def\@tempb {arXiv}\fi \@ifundefined
  {mn@eprint@\@tempb}{\@tempb:\@tempc}{\expandafter \expandafter \csname
  mn@eprint@\@tempb\endcsname \expandafter{\@tempc}}}

\bibitem[\protect\citeauthoryear{{Abazajian} et~al.,}{{Abazajian}
  et~al.}{2016}]{Abazajian2016}
{Abazajian} K.~N.,  et~al., 2016, preprint, \href
  {http://adsabs.harvard.edu/abs/2016arXiv161002743A} {} (\mn@eprint {arXiv}
  {1610.02743})

\bibitem[\protect\citeauthoryear{{Aghanim} et~al.,}{{Aghanim}
  et~al.}{2019}]{Aghanim2019}
{Aghanim} N.,  et~al., 2019, \mn@doi [\aap] {10.1051/0004-6361/201935271},
  \href {https://ui.adsabs.harvard.edu/abs/2019A&A...632A..47A} {632, A47}

\bibitem[\protect\citeauthoryear{{Allen}, {Evrard}  \& {Mantz}}{{Allen}
  et~al.}{2011}]{Allen2011}
{Allen} S.~W.,  {Evrard} A.~E.,   {Mantz} A.~B.,  2011, \mn@doi [\araa]
  {10.1146/annurev-astro-081710-102514}, \href
  {http://adsabs.harvard.edu/abs/2011ARA%26A..49..409A} {49, 409}

\bibitem[\protect\citeauthoryear{{Alonso}, {Sanchez}, {Slosar}  \& {LSST Dark
  Energy Science Collaboration}}{{Alonso} et~al.}{2019}]{Alonso2019}
{Alonso} D.,  {Sanchez} J.,  {Slosar} A.,   {LSST Dark Energy Science
  Collaboration} 2019, \mn@doi [\mnras] {10.1093/mnras/stz093}, \href
  {https://ui.adsabs.harvard.edu/abs/2019MNRAS.484.4127A} {484, 4127}

\bibitem[\protect\citeauthoryear{{Arnaud}, {Pratt}, {Piffaretti},
  {B{\"o}hringer}, {Croston}  \& {Pointecouteau}}{{Arnaud}
  et~al.}{2010}]{Arnaud2010}
{Arnaud} M.,  {Pratt} G.~W.,  {Piffaretti} R.,  {B{\"o}hringer} H.,  {Croston}
  J.~H.,   {Pointecouteau} E.,  2010, \mn@doi [\aap]
  {10.1051/0004-6361/200913416}, \href
  {http://adsabs.harvard.edu/abs/2010A%26A...517A..92A} {517, A92}

\bibitem[\protect\citeauthoryear{{Bleem}, {Stalder}, {de Haan}  et~al.}{{Bleem}
  et~al.}{2015}]{Bleem2015}
{Bleem} L.~E.,  {Stalder} B.,  {de Haan} T.,   et~al., 2015, \mn@doi [\apjs]
  {10.1088/0067-0049/216/2/27}, \href
  {http://adsabs.harvard.edu/abs/2015ApJS..216...27B} {216, 27}

\bibitem[\protect\citeauthoryear{{Bocquet}, {Dietrich}, {Schrabback}, {Bleem},
  {Klein}  et~al.}{{Bocquet} et~al.}{2019}]{Bocquet2018}
{Bocquet} S.,  {Dietrich} J.~P.,  {Schrabback} T.,  {Bleem} L.~E.,  {Klein} M.,
    et~al., 2019, \mn@doi [\apj] {10.3847/1538-4357/ab1f10}, \href
  {https://ui.adsabs.harvard.edu/abs/2019ApJ...878...55B} {878, 55}

\bibitem[\protect\citeauthoryear{Carlstrom, Holder  \& Reese}{Carlstrom
  et~al.}{2002}]{Carlstrom2002}
Carlstrom J.~E.,  Holder G.~P.,   Reese E.~D.,  2002, \mn@doi [Annual Review of
  Astronomy and Astrophysics] {10.1146/annurev.astro.40.060401.093803}, 40,
  643–680

\bibitem[\protect\citeauthoryear{{G{\'o}rski}, {Hivon}, {Banday}, {Wandelt},
  {Hansen}, {Reinecke}  \& {Bartelmann}}{{G{\'o}rski}
  et~al.}{2005}]{Gorski2005}
{G{\'o}rski} K.~M.,  {Hivon} E.,  {Banday} A.~J.,  {Wandelt} B.~D.,  {Hansen}
  F.~K.,  {Reinecke} M.,   {Bartelmann} M.,  2005, \mn@doi [\apj]
  {10.1086/427976}, \href {http://adsabs.harvard.edu/abs/2005ApJ...622..759G}
  {622, 759}

\bibitem[\protect\citeauthoryear{{Gruetjen}, {Fergusson}, {Liguori}  \&
  {Shellard}}{{Gruetjen} et~al.}{2017}]{Gruetjen2015}
{Gruetjen} H.~F.,  {Fergusson} J.~R.,  {Liguori} M.,   {Shellard} E.~P.~S.,
  2017, \mn@doi [\prd] {10.1103/PhysRevD.95.043532}, \href
  {http://adsabs.harvard.edu/abs/2017PhRvD..95d3532G} {95, 043532}

\bibitem[\protect\citeauthoryear{{Hasselfield}, {Hilton}, {Marriage}
  et~al.}{{Hasselfield} et~al.}{2013}]{Hasselfield2013}
{Hasselfield} M.,  {Hilton} M.,  {Marriage} T.~A.,   et~al., 2013, \mn@doi
  [\jcap] {10.1088/1475-7516/2013/07/008}, \href
  {http://adsabs.harvard.edu/abs/2013JCAP...07..008H} {7, 008}

\bibitem[\protect\citeauthoryear{{Herranz}, {Sanz}, {Hobson}, {Barreiro},
  {Diego}, {Mart{\'\i}nez-Gonz{\'a}lez}  \& {Lasenby}}{{Herranz}
  et~al.}{2002}]{Herranz2002}
{Herranz} D.,  {Sanz} J.~L.,  {Hobson} M.~P.,  {Barreiro} R.~B.,  {Diego}
  J.~M.,  {Mart{\'\i}nez-Gonz{\'a}lez} E.,   {Lasenby} A.~N.,  2002, \mn@doi
  [\mnras] {10.1046/j.1365-8711.2002.05704.x}, \href
  {https://ui.adsabs.harvard.edu/abs/2002MNRAS.336.1057H} {336, 1057}

\bibitem[\protect\citeauthoryear{{Hilton} et~al.,}{{Hilton}
  et~al.}{2021}]{Hilton2020}
{Hilton} M.,  et~al., 2021, \mn@doi [\apjs] {10.3847/1538-4365/abd023}, \href
  {https://ui.adsabs.harvard.edu/abs/2021ApJS..253....3H} {253, 3}

\bibitem[\protect\citeauthoryear{{Hivon}, {G{\'o}rski}, {Netterfield}, {Crill},
  {Prunet}  \& {Hansen}}{{Hivon} et~al.}{2002}]{Hivon2002}
{Hivon} E.,  {G{\'o}rski} K.~M.,  {Netterfield} C.~B.,  {Crill} B.~P.,
  {Prunet} S.,   {Hansen} F.,  2002, \mn@doi [\apj] {10.1086/338126}, \href
  {https://ui.adsabs.harvard.edu/abs/2002ApJ...567....2H} {567, 2}

\bibitem[\protect\citeauthoryear{{Melin}, {Bartlett}  \&
  {Delabrouille}}{{Melin} et~al.}{2006}]{Melin2006}
{Melin} J.~B.,  {Bartlett} J.~G.,   {Delabrouille} J.,  2006, \mn@doi [\aap]
  {10.1051/0004-6361:20065034}, \href
  {https://ui.adsabs.harvard.edu/abs/2006A&A...459..341M} {459, 341}

\bibitem[\protect\citeauthoryear{{Melin}, {Bartlett}, {Tarr{\'\i}o}  \&
  {Pratt}}{{Melin} et~al.}{2021}]{Melin2021}
{Melin} J.~B.,  {Bartlett} J.~G.,  {Tarr{\'\i}o} P.,   {Pratt} G.~W.,  2021,
  \mn@doi [\aap] {10.1051/0004-6361/202039471}, \href
  {https://ui.adsabs.harvard.edu/abs/2021A&A...647A.106M} {647, A106}

\bibitem[\protect\citeauthoryear{{Mroczkowski}, {Nagai}  et~al.}{{Mroczkowski}
  et~al.}{2019}]{Mroczkowski2019}
{Mroczkowski} T.,  {Nagai} D.,   et~al., 2019, \mn@doi [\ssr]
  {10.1007/s11214-019-0581-2}, \href
  {https://ui.adsabs.harvard.edu/abs/2019SSRv..215...17M} {215, 17}

\bibitem[\protect\citeauthoryear{{Planck 2013 results XX}}{{Planck 2013 results
  XX}}{2014}]{Planck2014XX}
{Planck 2013 results XX} 2014, \mn@doi [\aap] {10.1051/0004-6361/201321521},
  \href {https://ui.adsabs.harvard.edu/abs/2014A&A...571A..20P} {571, A20}

\bibitem[\protect\citeauthoryear{{Planck 2013 results XXIX}}{{Planck 2013
  results XXIX}}{2014}]{Planck2014}
{Planck 2013 results XXIX} 2014, \mn@doi [\aap] {10.1051/0004-6361/201321523},
  \href {http://adsabs.harvard.edu/abs/2014A%26A...571A..29P} {571, A29}

\bibitem[\protect\citeauthoryear{{Planck 2015 results VIII}}{{Planck 2015
  results VIII}}{2016}]{Planck2016viii}
{Planck 2015 results VIII} 2016, \mn@doi [\aap] {10.1051/0004-6361/201525820},
  \href {https://ui.adsabs.harvard.edu/abs/2016A&A...594A...8P} {594, A8}

\bibitem[\protect\citeauthoryear{{Planck 2015 results XXIV}}{{Planck 2015
  results XXIV}}{2016}]{Ade2016}
{Planck 2015 results XXIV} 2016, \mn@doi [\aap] {10.1051/0004-6361/201525833},
  \href {http://adsabs.harvard.edu/abs/2016A%26A...594A..24P} {594, A24}

\bibitem[\protect\citeauthoryear{{Planck 2015 results XXVII}}{{Planck 2015
  results XXVII}}{2016}]{Planck2016xxvii}
{Planck 2015 results XXVII} 2016, \mn@doi [\aap] {10.1051/0004-6361/201525823},
  \href {http://adsabs.harvard.edu/abs/2016A%26A...594A..27P} {594, A27}

\bibitem[\protect\citeauthoryear{{Pratt}, {Arnaud}, {Biviano}, {Eckert},
  {Ettori}, {Nagai}, {Okabe}  \& {Reiprich}}{{Pratt} et~al.}{2019}]{Pratt2019}
{Pratt} G.~W.,  {Arnaud} M.,  {Biviano} A.,  {Eckert} D.,  {Ettori} S.,
  {Nagai} D.,  {Okabe} N.,   {Reiprich} T.~H.,  2019, \mn@doi [\ssr]
  {10.1007/s11214-019-0591-0}, \href
  {http://adsabs.harvard.edu/abs/2019SSRv..215...25P} {215, 25}

\bibitem[\protect\citeauthoryear{{Salvati} et~al.,}{{Salvati}
  et~al.}{2021}]{Salvati2021}
{Salvati} L.,  et~al., 2021, arXiv e-prints, \href
  {https://ui.adsabs.harvard.edu/abs/2021arXiv211203606S} {p. arXiv:2112.03606}

\bibitem[\protect\citeauthoryear{{Simons Observatory Collaboration}}{{Simons
  Observatory Collaboration}}{2019}]{SO2019}
{Simons Observatory Collaboration} 2019, \mn@doi [\jcap]
  {10.1088/1475-7516/2019/02/056}, \href
  {https://ui.adsabs.harvard.edu/abs/2019JCAP...02..056A} {2019, 056}

\bibitem[\protect\citeauthoryear{{Stein}, {Alvarez}  \& {Bond}}{{Stein}
  et~al.}{2019}]{Stein2019}
{Stein} G.,  {Alvarez} M.~A.,   {Bond} J.~R.,  2019, \mn@doi [\mnras]
  {10.1093/mnras/sty3226}, \href
  {https://ui.adsabs.harvard.edu/abs/2019MNRAS.483.2236S} {483, 2236}

\bibitem[\protect\citeauthoryear{{Stein}, {Alvarez}, {Bond}, {van Engelen}  \&
  {Battaglia}}{{Stein} et~al.}{2020}]{Stein2020}
{Stein} G.,  {Alvarez} M.~A.,  {Bond} J.~R.,  {van Engelen} A.,   {Battaglia}
  N.,  2020, \mn@doi [\jcap] {10.1088/1475-7516/2020/10/012}, \href
  {https://ui.adsabs.harvard.edu/abs/2020JCAP...10..012S} {2020, 012}

\bibitem[\protect\citeauthoryear{{Sunyaev} \& {Zeldovich}}{{Sunyaev} \&
  {Zeldovich}}{1972}]{Sunyaev1972}
{Sunyaev} R.~A.,  {Zeldovich} Y.~B.,  1972, Comments on Astrophysics and Space
  Physics, \href {http://adsabs.harvard.edu/abs/1972CoASP...4..173S} {4, 173}

\bibitem[\protect\citeauthoryear{{Tarr{\'\i}o}, {Melin}  \&
  {Arnaud}}{{Tarr{\'\i}o} et~al.}{2019}]{Tarrio2019}
{Tarr{\'\i}o} P.,  {Melin} J.~B.,   {Arnaud} M.,  2019, \mn@doi [\aap]
  {10.1051/0004-6361/201834979}, \href
  {https://ui.adsabs.harvard.edu/abs/2019A&A...626A...7T} {626, A7}

\bibitem[\protect\citeauthoryear{{Vanderlinde}, {Crawford}, {de Haan},
  {Dudley}, {Shaw}  et~al.}{{Vanderlinde} et~al.}{2010}]{Vanderlinde2010}
{Vanderlinde} K.,  {Crawford} T.~M.,  {de Haan} T.,  {Dudley} J.~P.,  {Shaw}
  L.,   et~al., 2010, \mn@doi [\apj] {10.1088/0004-637X/722/2/1180}, \href
  {https://ui.adsabs.harvard.edu/abs/2010ApJ...722.1180V} {722, 1180}

\bibitem[\protect\citeauthoryear{{Zubeldia} \& {Challinor}}{{Zubeldia} \&
  {Challinor}}{2019}]{Zubeldia2019}
{Zubeldia} {\'{I}}.,  {Challinor} A.,  2019, \mn@doi [\mnras]
  {10.1093/mnras/stz2153}, \href
  {https://ui.adsabs.harvard.edu/abs/2019MNRAS.489..401Z} {489, 401}

\bibitem[\protect\citeauthoryear{{Zubeldia}, {Rotti}, {Chluba}  \&
  {Battye}}{{Zubeldia} et~al.}{2021}]{Zubeldia2021}
{Zubeldia} {\'I}.,  {Rotti} A.,  {Chluba} J.,   {Battye} R.,  2021, \mn@doi
  [\mnras] {10.1093/mnras/stab2461}, \href
  {https://ui.adsabs.harvard.edu/abs/2021MNRAS.tmp.2252Z} {}

\makeatother
\end{thebibliography}




\appendix

\section{Calculation of the matched filter covariance bias}\label{appendix}

As noted in Section \ref{sec:mmf}, the MMF signal-to-noise can be written as

\begin{equation}
    q = N(\mathbfit{n})^{-1/2} \int \frac{d^2 \mathbfit{l}}{2 \pi}  \mathbfit{y}_{\mathrm{t}}^{\dagger} (\mathbfit{l})   \tilde{\mathbfss{C}}^{-1} (l ; \mathbfit{n}) \mathbfit{d} (\mathbfit{l}),
\end{equation}
where $\mathbfit{y}_{\mathrm{t}}(\mathbfit{l})$ is the MMF template; $\mathbfit{d} (\mathbfit{l})$ is the data, which can in turn be written as $\mathbfit{d} (\mathbfit{l}) = \mathbfit{y} (\mathbfit{l}) + \mathbfit{n} (\mathbfit{l})$, where  $\mathbfit{y} (\mathbfit{l})$ is the signal and $\mathbfit{n} (\mathbfit{l})$, the noise (which can also include foregrounds); and $\tilde{\mathbfss{C}} (l ; \mathbfit{n})$ is the \textit{data} covariance matrix estimate, which we assume to be homogeneous and isotropic. The normalisation $N(\mathbfit{n})$ can be written as

\begin{equation}
    N(\mathbfit{n}) = \int \frac{d^2 \mathbfit{l}}{2 \pi}  \mathbfit{y}_{\mathrm{t}}^{\dagger} (\mathbfit{l})   \tilde{\mathbfss{C}}^{-1} (l ; \mathbfit{n}) \mathbfit{y}_{\mathrm{t}} (\mathbfit{l}).
\end{equation}
We have assume that the covariance matrix $\tilde{\mathbfss{C}} (l ; \mathbfit{n})$ is estimated from the data. Thus, it depends on the noise realisation $\mathbfit{n} (\mathbfit{l})$; as a consequence, so does the normalisation $N(\mathbfit{n})$. We take $\tilde{\mathbfss{C}} (l ; \mathbfit{n})$ to be the the cross-frequency power spectra of the data averaged over some multipole annuli,

\begin{multline}\label{covarianceestimate}
\tilde{\mathbfss{C}}_{ij} (l ; \mathbfit{n})  = V(l)^{-1} \int_{l \, \mathrm{bin}}  \frac{d^2 \mathbfit{l}^{\prime}}{2 \pi} d_i^{\dagger}(\mathbfit{l}^{\prime}) d_j(\mathbfit{l}^{\prime}) 
\\
= V(l)^{-1} \int_{l \, \mathrm{bin}} \frac{d^2 \mathbfit{l}^{\prime}}{2 \pi} [ y_i^{\ast}(\mathbfit{l}^{\prime}) y_j(\mathbfit{l}^{\prime}) + 
    y_i^{\ast}(\mathbfit{l}^{\prime}) n_j(\mathbfit{l}^{\prime}) + 
    n_i^{\ast}(\mathbfit{l}^{\prime}) y_j(\mathbfit{l}^{\prime})   \\ +
    n_i^{\ast}(\mathbfit{l}^{\prime}) n_j(\mathbfit{l}^{\prime}) ] ,
\end{multline}
where the integral is performed within the annulus (or bin) containing multipole $l$, and where $V(l)$ is the area of the annulus. The tilde above the covariance matrix symbol denotes that it is averaged over some bins. We note that this notation is not used in the rest of the paper, where there is no confusion with unbinned covariance matrices.

We want to compute the covariance bias of Section \ref{subsec:ps}, $ \left\langle q - \bar{q}_{\mathrm{t}} \right \rangle$. This can be written as

\begin{equation}\label{mmfbias}
     \left\langle q - \bar{q}_{\mathrm{t}} \right \rangle = \langle N(\mathbfit{n})^{-1/2} \int \frac{ d^2 \mathbfit{l}}{2 \pi}  \mathbfit{y}_{\mathrm{t}}^{\dagger} (\mathbfit{l})   \tilde{\mathbfss{C}}^{-1} (l ; \mathbfit{n}) \mathbfit{n} (\mathbfit{l}) \rangle,
\end{equation}
where angular brackets denote ensemble averaging over noise realisations. We do so by expanding this expression to lowest order in the covariance perturbation, which we define as $\Delta  \tilde{\mathbfss{C}} (l ; \mathbfit{d}) \equiv \tilde{\mathbfss{C}} (l ; \mathbfit{d}) - \tilde{\mathbfss{C}}_0 (l) $, where $ \tilde{\mathbfss{C}}_0 (l)$ is the true mean covariance, which does not depend on the noise realisation. Two terms arise: one from the integral in Eq. (\ref{mmfbias}), and another one from the normalisation $N(\mathbfit{n})$. We can thus write  $\left\langle q - \bar{q}_{\mathrm{t}} \right \rangle \equiv \Delta \bar{q}_1 +  \Delta \bar{q}_2$. To lowest (linear) order in the covariance perturbation, the inverse covariance can be written as 

\begin{equation}
     \tilde{\mathbfss{C}}^{-1} (l ; \mathbfit{n}) = \left[ 1 - \tilde{\mathbfss{C}}_0^{-1} (l)  \Delta \tilde{\mathbfss{C}} (l ; \mathbfit{n})  \right] \tilde{\mathbfss{C}}_0^{-1}.
\end{equation}
The first term in the covariance bias $\Delta \bar{q}_1$, coming from the integral in Eq. (\ref{mmfbias}), is then given by

\begin{equation}
    \Delta \bar{q}_1 = -N_0^{-1/2} \int \frac{ d^2 \mathbfit{l}}{2 \pi} \,  \mathbfit{y}_{\mathrm{t}}^{\dagger} (\mathbfit{l})   \tilde{\mathbfss{C}}_0^{-1} (l) \langle  \Delta \tilde{\mathbfss{C}} (l ; \mathbfit{n})  \tilde{\mathbfss{C}}_0^{-1} (l)  \mathbfit{n}(\mathbfit{l}) \rangle,
\end{equation}
where $N_0$ is the normalisation evaluated with the true, noiseless data covariance $\tilde{\mathbfss{C}}_0$. Substituting the expression for the covariance estimate, given by Eq. (\ref{covarianceestimate}), assuming the noise to be statistically homogeneous and isotropic, and neglecting the noise bispectrum, yields

\begin{equation}\label{bias1}
    \Delta \bar{q}_1 = -N_0^{-1/2}\int  \frac{ d^2 \mathbfit{l}}{2 \pi} \,  \mathbfit{y}_{\mathrm{t}}^{\dagger} (\mathbfit{l})   \tilde{\mathbfss{C}}_0^{-1} (l)  \mathbfit{b}_{1} (\mathbfit{l}),
\end{equation}
where
\begin{equation}\label{b1}
     \mathbfit{b}_{1} (\mathbfit{l}) =   V(l)^{-1} [ \mathbfss{N} (l) \, \tilde{\mathbfss{C}}_0^{-1} (l) \mathbfit{y} (\mathbfit{l}) + \mathrm{Tr} [ \tilde{\mathbfss{C}}_0^{-1}  (l)  \mathbfss{N} (l) ] \mathbfit{y}^{\dagger}(\mathbfit{l})],
\end{equation}
where $\mathbfss{N} (l)$ is the \textit{unbinned} \textit{noise} covariance matrix. Notice that, to this order in the covariance perturbation, the bias vanishes if the signal is not present in the covariance matrix estimate, as $\mathbfit{y} (\mathbfit{l})$ and  $\mathbfit{y}^{\dagger} (\mathbfit{l})$ in Eq. (\ref{b1}) arise due to their presence in the covariance estimate (see Eq. \ref{covarianceestimate}).

In order to compute the second term in the covariance bias, $\Delta \bar{q}_2$, we write the normalisation as $N(\mathbfit{d}) = N_0 + \Delta N(\mathbfit{n})$. Then, to lowest order, the square root of its inverse is given by

\begin{equation}
    N(\mathbfit{d})^{-1/2} = N_0^{-1/2} \left[ 1 - \frac{\Delta N (\mathbfit{n})}{2 N_0} \right],
\end{equation}
where $\Delta N (\mathbfit{n})$ is the normalisation perturbation, and hence,

\begin{equation}\label{q2calc}
    \Delta \bar{q}_2 = \frac{1}{2} N_0^{-3/2} \int  \frac{ d^2 \mathbfit{l}}{2 \pi} \mathbfit{y}_{\mathrm{t}}^{\dagger} (\mathbfit{l})   \tilde{\mathbfss{C}}_0^{-1} (l) \langle \Delta N (\mathbfit{n}) \mathbfit{n}  (\mathbfit{l})\rangle.
\end{equation}
The normalisation perturbation $\Delta N(\mathbfit{n})$ is calculated by expanding the normalisation to lowest order in the covariance perturbation,

\begin{equation}\label{normpert}
    \Delta N(\mathbfit{n}) = \int  \frac{ d^2 \mathbfit{l}}{2 \pi}\, \mathbfit{y}_{\mathrm{t}}^{\dagger} (\mathbfit{l})   \tilde{\mathbfss{C}}_0^{-1} (l) \Delta \tilde{\mathbfss{C}} (l ; \mathbfit{n})  \tilde{\mathbfss{C}}_0^{-1} (l)   \mathbfit{y}_{\mathrm{t}} (\mathbfit{l}).
\end{equation}
In an analogous calculation to that of the first term in the covariance bias, plugging in the expression for the covariance estimate, given by Eq. (\ref{covarianceestimate}), and for the normalisation perturbation, given by Eq. (\ref{normpert}), into Eq. (\ref{q2calc}), and neglecting the noise bispectrum, $\Delta \bar{q}_2 $ reads

\begin{equation}\label{bias2}
    \Delta \bar{q}_2 = \frac{1}{2}  N_0^{-3/2}   \int  \frac{ d^2 \mathbfit{l}}{2 \pi} \mathbfit{y}_{\mathrm{t}}^{\dagger} (\mathbfit{l})   \tilde{\mathbfss{C}}_0^{-1} (l)  \mathbfit{b}_{2} (\mathbfit{l}),
\end{equation}
where
\begin{equation}
    \mathbfit{b}_{2} (\mathbfit{l}) =  V(l)^{-1}  \mathcal{R} \{ [ \mathbfit{y}^{\mathrm{\dagger}} (\mathbfit{l})  \tilde{\mathbfss{C}}_0^{-1} (l) \mathbfit{y} (\mathbfit{l}) ]  \mathbfss{N} (l)  \tilde{\mathbfss{C}}_0^{-1} (l)  \mathbfit{y} (\mathbfit{l}) \},
\end{equation}
where $\mathcal{R} \{z\}$ denotes the real part of $z$. As it was the case for $\Delta \bar{q}_1$, $\Delta \bar{q}_2 $ also vanishes to this order in the covariance perturbation if the signal $ \mathbfit{y}(\mathbfit{l})$ is not present in the covariance estimate.

The covariance bias on the MMF amplitude estimator, $\hat{y}_0$, can be calculated in an analogous fashion. As expected, two terms arise as well. The first term is given by Eq. (\ref{bias1}) substituting $N_0^{-1/2}$ by $N_0^{-1}$, whereas the second term is given by Eq. (\ref{bias2}) substituting $N_0^{-3/2}/2$ by $N_0^{-2}$. As was the case for the signal-to-noise, the bias vanishes if the signal is not present in the covariance estimate.


\bsp	
\label{lastpage}
\end{document}